\pdfoutput=1

\documentclass[12pt,a4paper]{article}

\usepackage{ifthen} 
\usepackage{overpic}
\newboolean{pdflatex}
\setboolean{pdflatex}{true} 

\newboolean{articletitles}
\setboolean{articletitles}{true} 

\newboolean{uprightparticles}
\setboolean{uprightparticles}{false} 

\usepackage{microtype}
\usepackage{lineno}  
\usepackage{xspace} 
\usepackage{caption} 

\usepackage{graphicx}  
\usepackage{color}
\usepackage{colortbl}
\graphicspath{{./figs/}} 

\usepackage{amsmath} 
\usepackage{amssymb}
\usepackage{amsfonts}
\usepackage{upgreek} 

\newcommand*\patchAmsMathEnvironmentForLineno[1]{%
\expandafter\let\csname old#1\expandafter\endcsname\csname #1\endcsname
\expandafter\let\csname oldend#1\expandafter\endcsname\csname
end#1\endcsname
 \renewenvironment{#1}%
   {\linenomath\csname old#1\endcsname}%
   {\csname oldend#1\endcsname\endlinenomath}%
}
\newcommand*\patchBothAmsMathEnvironmentsForLineno[1]{%
  \patchAmsMathEnvironmentForLineno{#1}%
  \patchAmsMathEnvironmentForLineno{#1*}%
}
\AtBeginDocument{%
\patchBothAmsMathEnvironmentsForLineno{equation}%
\patchBothAmsMathEnvironmentsForLineno{align}%
\patchBothAmsMathEnvironmentsForLineno{flalign}%
\patchBothAmsMathEnvironmentsForLineno{alignat}%
\patchBothAmsMathEnvironmentsForLineno{gather}%
\patchBothAmsMathEnvironmentsForLineno{multline}%
}

\usepackage{hyperref}    
\usepackage[all]{hypcap} 

\def\lhcb {\mbox{LHCb}\xspace}

\ifthenelse{\boolean{uprightparticles}}%
{

 \def\Pmu         {\ensuremath{\upmu}\xspace}

 \def\Ppi         {\ensuremath{\uppi}\xspace}

 \def\Ppsi        {\ensuremath{\uppsi}\xspace}

 \def\PDelta      {\ensuremath{\Delta}\xspace}                 
 \def\PXi      {\ensuremath{\Xi}\xspace}                 
 \def\PLambda      {\ensuremath{\Lambda}\xspace}                 
 \def\PSigma      {\ensuremath{\Sigma}\xspace}                 
 \def\POmega      {\ensuremath{\Omega}\xspace}                 
 \def\PUpsilon      {\ensuremath{\Upsilon}\xspace}                 
 
 \def\PB      {\ensuremath{\mathrm{B}}\xspace}                 
                  
 \def\PD      {\ensuremath{\mathrm{D}}\xspace}

 \def\PJ      {\ensuremath{\mathrm{J}}\xspace}                 
 \def\PK      {\ensuremath{\mathrm{K}}\xspace}

 \def\Pb      {\ensuremath{\mathrm{b}}\xspace}                 
 \def\Pc      {\ensuremath{\mathrm{c}}\xspace}

 \def\Pi      {\ensuremath{\mathrm{i}}\xspace}

}
{

 \def\Pmu         {\ensuremath{\mu}\xspace}

 \def\Ppi         {\ensuremath{\pi}\xspace}

 \def\Ppsi        {\ensuremath{\psi}\xspace}                 
                  
 \mathchardef\PDelta="7101
 \mathchardef\PXi="7104
 \mathchardef\PLambda="7103
 \mathchardef\PSigma="7106
 \mathchardef\POmega="710A
 \mathchardef\PUpsilon="7107
                  
 \def\PB      {\ensuremath{B}\xspace}                 
                  
 \def\PD      {\ensuremath{D}\xspace}

 \def\PJ      {\ensuremath{J}\xspace}                 
 \def\PK      {\ensuremath{K}\xspace}

 \def\Pb      {\ensuremath{b}\xspace}                 
 \def\Pc      {\ensuremath{c}\xspace}

 \def\Pi      {\ensuremath{i}\xspace}

}

\def\muon       {{\ensuremath{\Pmu}}\xspace}
\def\mup        {{\ensuremath{\Pmu^+}}\xspace}
\def\mun        {{\ensuremath{\Pmu^-}}\xspace} 
\def\mumu       {{\ensuremath{\Pmu^+\Pmu^-}}\xspace}

\def\cquark    {{\ensuremath{\Pc}}\xspace}

\def\bquark    {{\ensuremath{\Pb}}\xspace}

\def\pion   {{\ensuremath{\Ppi}}\xspace}

\def\pim    {{\ensuremath{\pion^-}}\xspace}

\def\kaon    {{\ensuremath{\PK}}\xspace}
  \def\Kbar    {{\kern 0.2em\overline{\kern -0.2em \PK}{}}\xspace}

\def\Kp      {{\ensuremath{\kaon^+}}\xspace}

\def\KS      {{\ensuremath{\kaon^0_{\rm\scriptscriptstyle S}}}\xspace}

\def\Kstarz  {{\ensuremath{\kaon^{*0}}}\xspace}

  \def\Dbar    {{\kern 0.2em\overline{\kern -0.2em \PD}{}}\xspace}
\def\D       {{\ensuremath{\PD}}\xspace}

\def\B       {{\ensuremath{\PB}}\xspace}
\def\Bbar    {{\ensuremath{\kern 0.18em\overline{\kern -0.18em \PB}{}}}\xspace}

\def\Bz      {{\ensuremath{\B^0}}\xspace}

\def\jpsi     {{\ensuremath{{\PJ\mskip -3mu/\mskip -2mu\Ppsi\mskip 2mu}}}\xspace}

  \def\Y#1S{\ensuremath{\PUpsilon{(#1S)}}\xspace}

\def\Lbar        {{\ensuremath{\kern 0.1em\overline{\kern -0.1em\PLambda}}}\xspace}

\newcommand{\decay}[2]{\ensuremath{#1\!\to #2}\xspace}         

\def\to                 {\ensuremath{\rightarrow}\xspace}

\def\AT#1     {\ensuremath{A_{\mathrm{T}}^{#1}}\xspace}           

\def\C#1      {\ensuremath{\mathcal{C}_{#1}}\xspace}                       
\def\Cp#1     {\ensuremath{\mathcal{C}_{#1}^{'}}\xspace}                    
\def\Ceff#1   {\ensuremath{\mathcal{C}_{#1}^{\mathrm{(eff)}}}\xspace}        
\def\Cpeff#1  {\ensuremath{\mathcal{C}_{#1}^{'\mathrm{(eff)}}}\xspace}       
\def\Ope#1    {\ensuremath{\mathcal{O}_{#1}}\xspace}                       
\def\Opep#1   {\ensuremath{\mathcal{O}_{#1}^{'}}\xspace}                    

\newcommand{\unit}[1]{\ensuremath{\rm\,#1}\xspace}          

\newcommand{\tev}{\ensuremath{\mathrm{\,Te\kern -0.1em V}}\xspace}
\newcommand{\gev}{\ensuremath{\mathrm{\,Ge\kern -0.1em V}}\xspace}
\newcommand{\mev}{\ensuremath{\mathrm{\,Me\kern -0.1em V}}\xspace}
\newcommand{\kev}{\ensuremath{\mathrm{\,ke\kern -0.1em V}}\xspace}
\newcommand{\ev}{\ensuremath{\mathrm{\,e\kern -0.1em V}}\xspace}
\newcommand{\gevc}{\ensuremath{{\mathrm{\,Ge\kern -0.1em V\!/}c}}\xspace}
\newcommand{\mevc}{\ensuremath{{\mathrm{\,Me\kern -0.1em V\!/}c}}\xspace}
\newcommand{\gevcc}{\ensuremath{{\mathrm{\,Ge\kern -0.1em V\!/}c^2}}\xspace}
\newcommand{\gevgevcccc}{\ensuremath{{\mathrm{\,Ge\kern -0.1em V^2\!/}c^4}}\xspace}
\newcommand{\mevcc}{\ensuremath{{\mathrm{\,Me\kern -0.1em V\!/}c^2}}\xspace}

\def\mm   {\ensuremath{\rm \,mm}\xspace}

\def\mum  {\ensuremath{{\,\upmu\rm m}}\xspace}

\def\invpb {\ensuremath{\mbox{\,pb}^{-1}}\xspace}

\def\invfb   {\ensuremath{\mbox{\,fb}^{-1}}\xspace}

\newcommand{\chisq}{\ensuremath{\chi^2}\xspace}
\newcommand{\chisqndf}{\ensuremath{\chi^2/\mathrm{ndf}}\xspace}
\newcommand{\chisqip}{\ensuremath{\chi^2_{\rm IP}}\xspace}
\newcommand{\ipchisq}{\ensuremath{\chi^2_{\rm IP}}\xspace}
\newcommand{\ip}{\ensuremath{{\rm IP}}\xspace}

\def\gsim{{~\raise.15em\hbox{$>$}\kern-.85em
          \lower.35em\hbox{$\sim$}~}\xspace}
\def\lsim{{~\raise.15em\hbox{$<$}\kern-.85em
          \lower.35em\hbox{$\sim$}~}\xspace}

\def\sqs   {\ensuremath{\protect\sqrt{s}}\xspace}

\def\ptot       {\mbox{$p$}\xspace}
\def\pt         {\mbox{$p_{\rm T}$}\xspace}

\def\dllmupi    {\ensuremath{\mathrm{DLL}_{\muon\pi}}\xspace}

\def\evtgen     {\mbox{\textsc{EvtGen}}\xspace}

\def\geant      {\mbox{\textsc{Geant4}}\xspace}

\def\photos     {\mbox{\textsc{Photos}}\xspace}

\def\pythia     {\mbox{\textsc{Pythia}}\xspace}

\def\tell1  {TELL1\xspace}
\def\ukl1   {UKL1\xspace}

\def\cfourften     {\ensuremath{\rm C_4 F_{10}}\xspace}

\usepackage{cite} 
\usepackage{mciteplus}

\usepackage{longtable} 

\begin{document}

\renewcommand{\thefootnote}{\fnsymbol{footnote}}
\setcounter{footnote}{1}


\begin{titlepage}
\pagenumbering{roman}

\vspace*{-1.5cm}
\centerline{\large EUROPEAN ORGANIZATION FOR NUCLEAR RESEARCH (CERN)}
\vspace*{1.5cm}
\hspace*{-0.5cm}
\begin{tabular*}{\linewidth}{lc@{\extracolsep{\fill}}r}
\ifthenelse{\boolean{pdflatex}}
{\vspace*{-2.7cm}\mbox{\!\!\!\includegraphics[width=.14\textwidth]{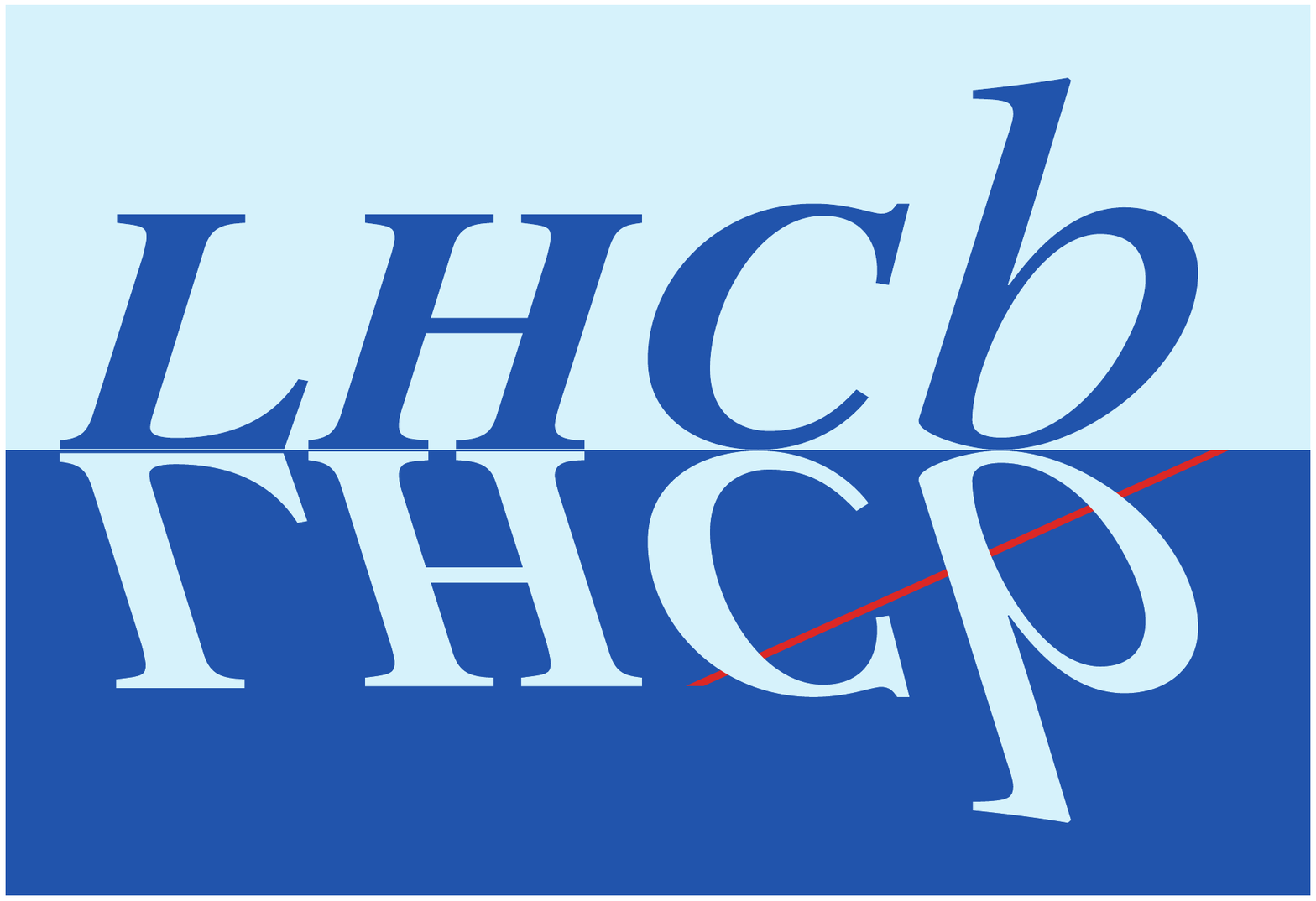}} & &}%
{\vspace*{-1.2cm}\mbox{\!\!\!\includegraphics[width=.12\textwidth]{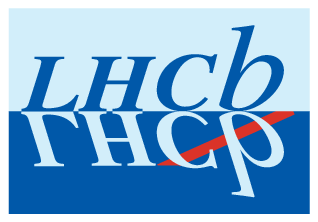}} & &}%
\\
 & & CERN-LHCb-DP-2013-002 \\
 & & February 13, 2015 \\ 
 & & \\
\end{tabular*}

\vspace*{4.0cm}

{\bf\boldmath\huge
\begin{center}
  Measurement of the track reconstruction efficiency at LHCb
\end{center}
}

\vspace*{2.0cm}

\begin{center}
The LHCb collaboration\footnote{Authors are listed on the following pages.}
\end{center}

\vspace{\fill}

\begin{abstract}
  \noindent

  The determination of track reconstruction efficiencies at LHCb using
  $\jpsi\to\mup\mun$ decays is presented. Efficiencies
  above $95\%$ are found for the data taking periods in 2010, 2011, and 2012.
  The ratio of the track reconstruction efficiency of muons in data and
  simulation is compatible with unity and measured with an uncertainty of $0.8\,\%$ for data taking in 2010,
  and at a precision of $0.4\,\%$ for data taking in 2011 and 2012. For hadrons an additional $1.4\,\%$ uncertainty due to material interactions is assumed. This result is crucial
  for accurate cross section and branching fraction measurements in \lhcb.

\end{abstract}

\vspace*{2.0cm}

\begin{center}
  Published in JINST 10 (2015) P02007
\end{center}

\vspace{\fill}

{\footnotesize 
\centerline{\copyright~CERN on behalf of the \lhcb collaboration, license \href{http://creativecommons.org/licenses/by/3.0/}{CC-BY-3.0}.}}
\vspace*{2mm}

\end{titlepage}


\newpage
\setcounter{page}{2}
\mbox{~}
\newpage

\centerline{\large\bf LHCb collaboration}
\begin{flushleft}
\small
R.~Aaij$^{41}$, 
B.~Adeva$^{37}$, 
M.~Adinolfi$^{46}$, 
A.~Affolder$^{52}$, 
Z.~Ajaltouni$^{5}$, 
S.~Akar$^{6}$, 
J.~Albrecht$^{9}$, 
F.~Alessio$^{38}$, 
M.~Alexander$^{51}$, 
S.~Ali$^{41}$
% ...

{\footnotesize \it
% ...
$ ^{5}$Clermont Universit\'{e}, Universit\'{e} Blaise Pascal, CNRS/IN2P3, LPC, Clermont-Ferrand, France\\
$ ^{6}$CPPM, Aix-Marseille Universit\'{e}, CNRS/IN2P3, Marseille, France\\
% ...
$ ^{9}$Fakult\"{a}t Physik, Technische Universit\"{a}t Dortmund, Dortmund, Germany\\
% ...
$ ^{37}$Universidad de Santiago de Compostela, Santiago de Compostela, Spain\\
$ ^{38}$European Organization for Nuclear Research (CERN), Geneva, Switzerland\\
% ...
$ ^{41}$Nikhef National Institute for Subatomic Physics, Amsterdam, The Netherlands\\
% ...
$ ^{46}$H.H. Wills Physics Laboratory, University of Bristol, Bristol, United Kingdom\\
% ...
$ ^{51}$School of Physics and Astronomy, University of Glasgow, Glasgow, United Kingdom\\
$ ^{52}$Oliver Lodge Laboratory, University of Liverpool, Liverpool, United Kingdom\\
% ...
}
\end{flushleft}

\cleardoublepage


\renewcommand{\thefootnote}{\arabic{footnote}}
\setcounter{footnote}{0}



\pagestyle{plain} 
\setcounter{page}{1}
\pagenumbering{arabic}


%

\section{Introduction}
\label{sec:Introduction}

The track reconstruction efficiency is an important quantity in
many physics analyses, especially those that aim at measuring a production cross
section or a branching fraction. 
The uncertainty on the track reconstruction efficiency was a source of large
systematic uncertainties with early LHCb data
\cite{LHCb-PAPER-2010-002}. The method presented in this paper has significantly reduced this uncertainty for recent measurements
\cite{LHCb-PAPER-2013-016}.

In physics analysis, the track reconstruction efficiency is usually estimated with simulated events.
To take possible differences between simulation and data into account, a data-driven correction procedure is applied. 
A clean sample of \decay{\jpsi}{\mumu} decays is selected in data with a tag-and-probe approach. \decay{\jpsi}{\mumu} decays are ideal candidates for efficiency measurements as they are abundant, clean, and the decay products cover the momentum spectrum needed in most physics analyses in LHCb. The purity of the sample is enhanced by selecting \jpsi from \bquark-hadron decays. The tag track is fully reconstructed and is well identified as a muon. The probe track is only partially reconstructed, not using information from at least one subdetector which is probed. The track reconstruction efficiency is determined by checking for the existence of a fully reconstructed track corresponding to the probe track as this allows to determine the efficiency of the subdetector that is not used in the reconstruction of the probe track.
It is calculated as a function of the
momentum of the probe track, $p$, its \mbox{pseudorapidity}, $\eta$, and the track
multiplicity of the event, $N_\text{track}$.
These are chosen because the efficiency is most affected by them. No strong dependence on the polar angle $\phi$ is observed.
The main result of this paper is the track reconstruction efficiency
ratio between data and simulation for prompt tracks and tracks from \B and \D mesons. This ratio is used in physics analyses to correct the track
reconstruction efficiency in simulated events and to determine its
uncertainty.
The measurement is performed on several
data samples to meet the requirements of the analyses performed at \lhcb. 
In this paper, the results are presented for the three data samples from run I, corresponding to different running conditions, proton-proton ($pp$) centre-of-mass energies and integrated luminosities: data taken in 2010 at \sqs = 7\tev corresponding to 29\invpb, data taken in 2011 at \sqs = 7\tev corresponding to 1\invfb, and data taken in 2012 at \sqs = 8\tev corresponding to 2\invfb. The 2010 results are valid for the full 2010 data set, corresponding to a luminosity of 37\invpb, since the same running conditions and track reconstruction were used throughout this period.

\section{Detector and software description}
\label{sec:Detector}

The \lhcb detector~\cite{LHCb-DP-2008-001} is a single-arm forward
spectrometer covering the \mbox{pseudorapidity} range $2<\eta <5$,
designed for the study of particles containing \bquark or \cquark
quarks. The detector includes a high-precision tracking system
consisting of a silicon-strip vertex detector,~VELO~\cite{VeloPerformance}, surrounding the $pp$
interaction region; a large-area silicon-strip detector,~TT~\cite{LHCb-TDR-009}, located
upstream of a dipole magnet with a bending power of about
$4{\rm\,Tm}$; and three stations of silicon-strip detectors (Inner Tracker)~\cite{LHCb-TDR-008} and straw
drift tubes (Outer Tracker)~\cite{LHCb-DP-2013-003} placed downstream of the magnet, called T stations.
The tracking system provides a measurement of momentum, \ptot,  with
a relative uncertainty that varies from 0.4\% at low momentum to 0.6\% at 100\gevc.
The minimum distance of a track to a primary vertex (PV), the impact parameter (\ip), is measured with a resolution of $(15+29/\pt)\mum$,
where \pt is the component of \ptot transverse to the beam, in \gevc.
The polarity
of the dipole magnet is reversed periodically throughout data taking. The
configuration with the magnetic field vertically upwards (downwards), bends
positively (negatively) charged particles in the horizontal plane towards the
centre of the LHC.
Different types of charged hadrons are distinguished using information
from two ring-imaging Cherenkov detectors, RICH1 and RICH2. Photon, electron, and
hadron candidates are identified by a calorimeter system consisting of
scintillating-pad and preshower detectors, an electromagnetic
calorimeter, and a hadronic calorimeter. Muons are identified by a
system composed of alternating layers of iron and multiwire
proportional chambers~\cite{LHCb-DP-2012-002}.
The trigger~\cite{LHCb-DP-2012-004} consists of a
hardware stage, based on information from the calorimeter and muon
systems, followed by a software stage, which applies a full event
reconstruction.

In the simulation, $pp$ collisions are generated using
\pythia~6.4~\cite{Sjostrand:2006za} with a specific \lhcb
configuration~\cite{LHCb-PROC-2010-056}.  Decays of hadronic particles
are described by \evtgen~\cite{Lange:2001uf}, in which final state
radiation is generated using \photos~\cite{Golonka:2005pn}. The
interaction of the generated particles with the detector and its
response are implemented using the \geant
toolkit~\cite{Allison:2006ve, *Agostinelli:2002hh} as described in
Ref.~\cite{LHCb-PROC-2011-006}. Hit inefficiencies, e.g. due to dead channels, are typically in the range 1-2\% and are included in the simulation. Differences in the positioning of the sensors between data and simulation are at the level of 0.5\mm. Both effects have a negligible impact on the tracking efficiency. The simulated events used in this study
are required to contain at least one $\decay{\jpsi}{\mup\mun}$ decay.

Differences in the response of the detectors in simulation and data could potentially lead to a different behaviour of the track reconstruction. The hit efficiencies have been measured in data using tracks. For the different subdetectors, they range from 98-100\%. Dead channels are included in the simulation, using an average over the data taking period. From simulations it is known that the (high) hit efficiency does not have any impact on the track reconstruction, as the algorithms have been written to be robust against small hit inefficiencies. The size of the sensitive detector elements are known very accurately and the positioning of the sensitive elements in the simulation is accurate at the level of 0.5\mm. Compared to the overall size of the tracking system, any inaccuracy at this level has negligible impact on the acceptance of the detector.

\section{Track reconstruction at LHCb}
\label{sec:generaltracking}

\begin{figure}
  \centering
  \includegraphics[width=.8\textwidth]{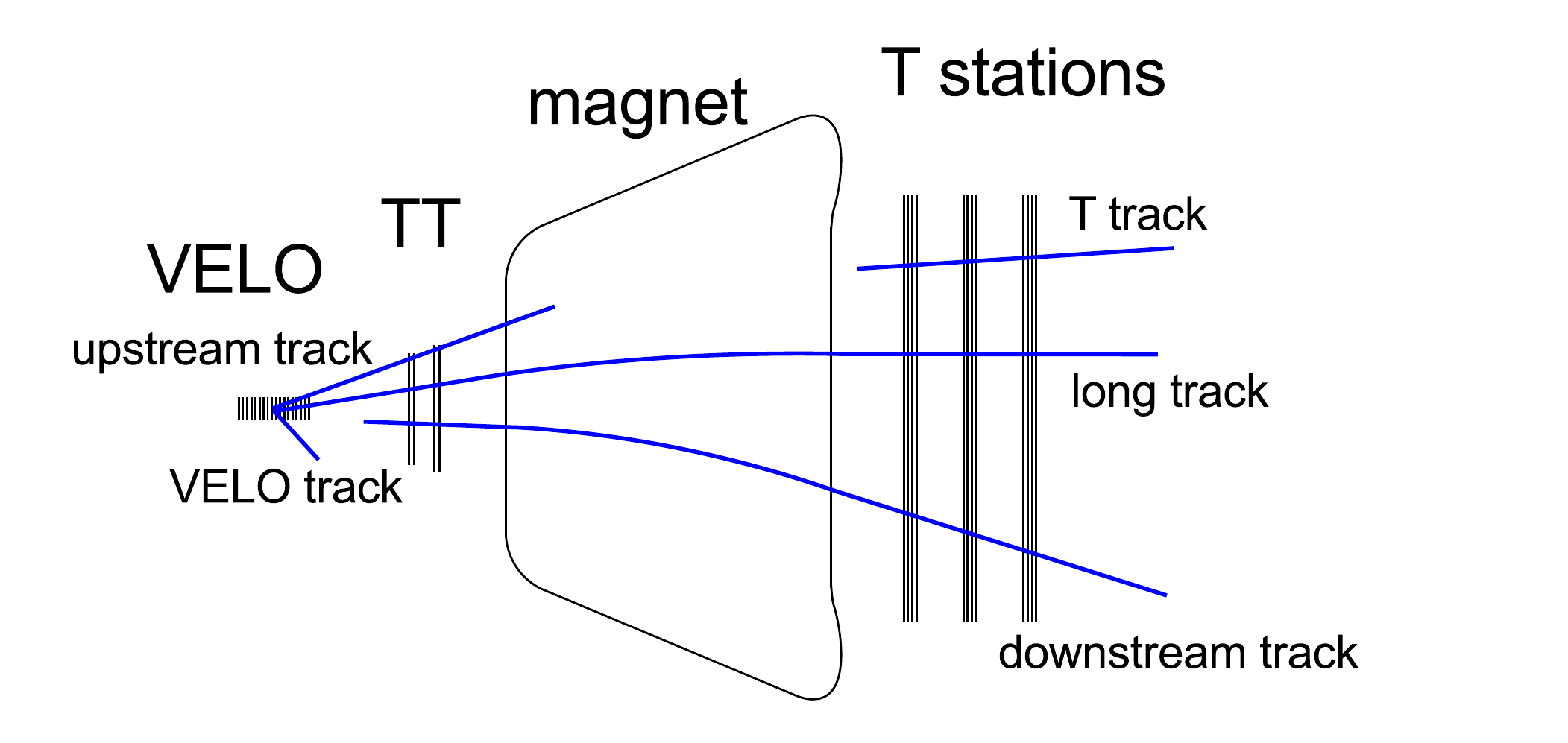}
  \caption{Tracking detectors and track types reconstructed by the track finding algorithms at LHCb.}
  \label{fig:ttypes}
\end{figure}

Owing to the design of the LHCb detector, which consists of tracking detectors mainly
outside the magnetic field, charged particle tracks are in approximation
straight line segments in the upstream part (VELO and TT) and in the downstream
part (T stations). Figure~\ref{fig:ttypes} shows an overview of the
different track types defined in the LHCb reconstruction: VELO tracks, which have hits
in the VELO; upstream tracks, which have hits in the two upstream trackers; T
tracks, which have hits in the T stations; downstream tracks, which have hits
in TT and the T stations; and long tracks, which have hits in the VELO and the T
stations. The latter tracks can additionally have hits in TT. 

If a particle is reconstructed more than once, as different track types, only the track best
suited for analysis purposes is kept. Hereby, long tracks are preferred over any
other track type, upstream tracks are preferred over VELO tracks, and downstream
tracks are preferred over T tracks. The number of unique tracks in an event,
$N_\text{track}$, is used in this study as a measure for the event
multiplicity; it is strongly correlated with the number of hits in the tracking detectors.
The number of tracks is chosen over the number of hits in a tracker to give a balanced measure of
the upstream and the downstream occupancy.

The reconstruction of long tracks starts with a search for VELO tracks\cite{Hutchcroft:2007ix}\cite{callot:2011a}. 
VELO tracks are reconstructed exploiting the fact that tracks form straight lines due to the absence of a magnetic field in the VELO. Two algorithms promote these VELO tracks to long tracks. The first algorithm,
called forward tracking\cite{callot:2007a}, combines VELO tracks with hits in the three T
stations. For a given VELO track and a single hit in one of the T stations the
momentum is fixed, enabling the algorithm to project hits in the T stations
along the trajectory. Hits which form clusters in the projection are used to
define the final long track. In the second algorithm, called track matching\cite{needham:2007a}\cite{Needham:2007zzb},
long tracks are made combining VELO tracks with T tracks, which are found by a
standalone track finding algorithm\cite{Callot:2008zza}.

If hits compatible with the long track trajectory are found in TT, they are added
to the track to improve the momentum resolution and as discrimination against fake tracks. This procedure is identical for the forward tracking and the track matching. 

Most analyses use long tracks because they provide the best momentum and spatial
resolution among all track types. Unless otherwise stated, track reconstruction
at LHCb refers to the reconstruction of long tracks. In a typical signal triggered event in 2011 or 2012, around 60 long tracks are reconstructed. Other track types, such as downstream tracks\cite{Callot:2007ii}, are used for the reconstruction of decay products of long-lived particles such as \KS mesons, or for internal alignment of the tracking detectors. They are reconstructed from T tracks, which are propagated back through the magnetic field to find corresponding hits in the TT stations.

The efficiency to
reconstruct charged particles as long tracks is determined in two
approaches. The first approach measures the track reconstruction
efficiency in the VELO and in the T stations individually and combines these efficiencies
to a single measurement. The second approach determines the efficiency to
reconstruct a long track directly.

\section{Tag-and-probe methods}
\label{sec:tagandprobe}

\begin{figure}
  \centering
  \begin{minipage}[b]{0.65\linewidth}
    \begin{overpic}[width=0.95\textwidth]{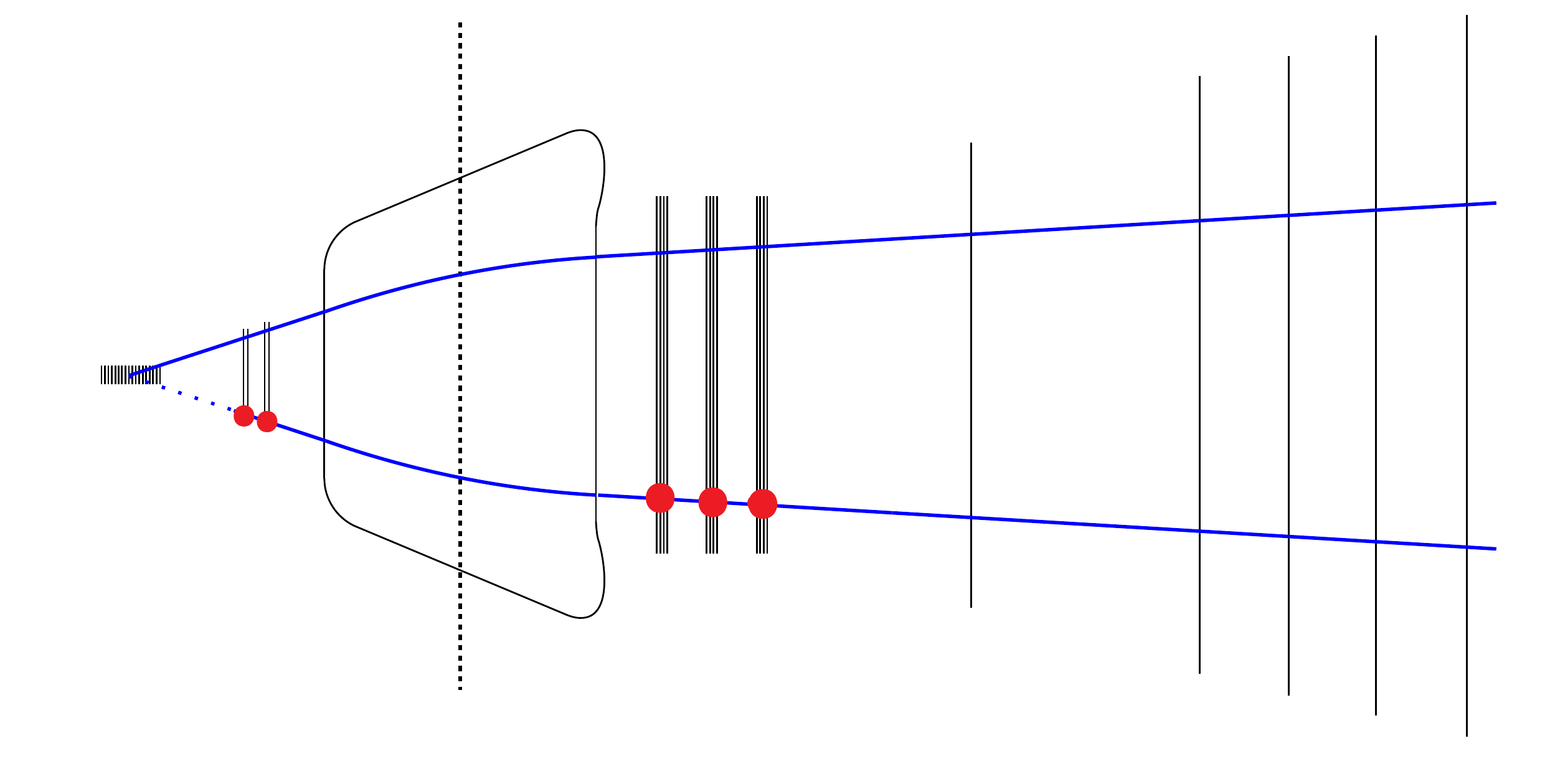}
      \put (5,40) {\small{(a)}}
    \end{overpic}
  \end{minipage}
  \hspace{.05\textwidth}
  \begin{minipage}[b]{0.65\linewidth}
    \begin{overpic}[width=0.95\textwidth]{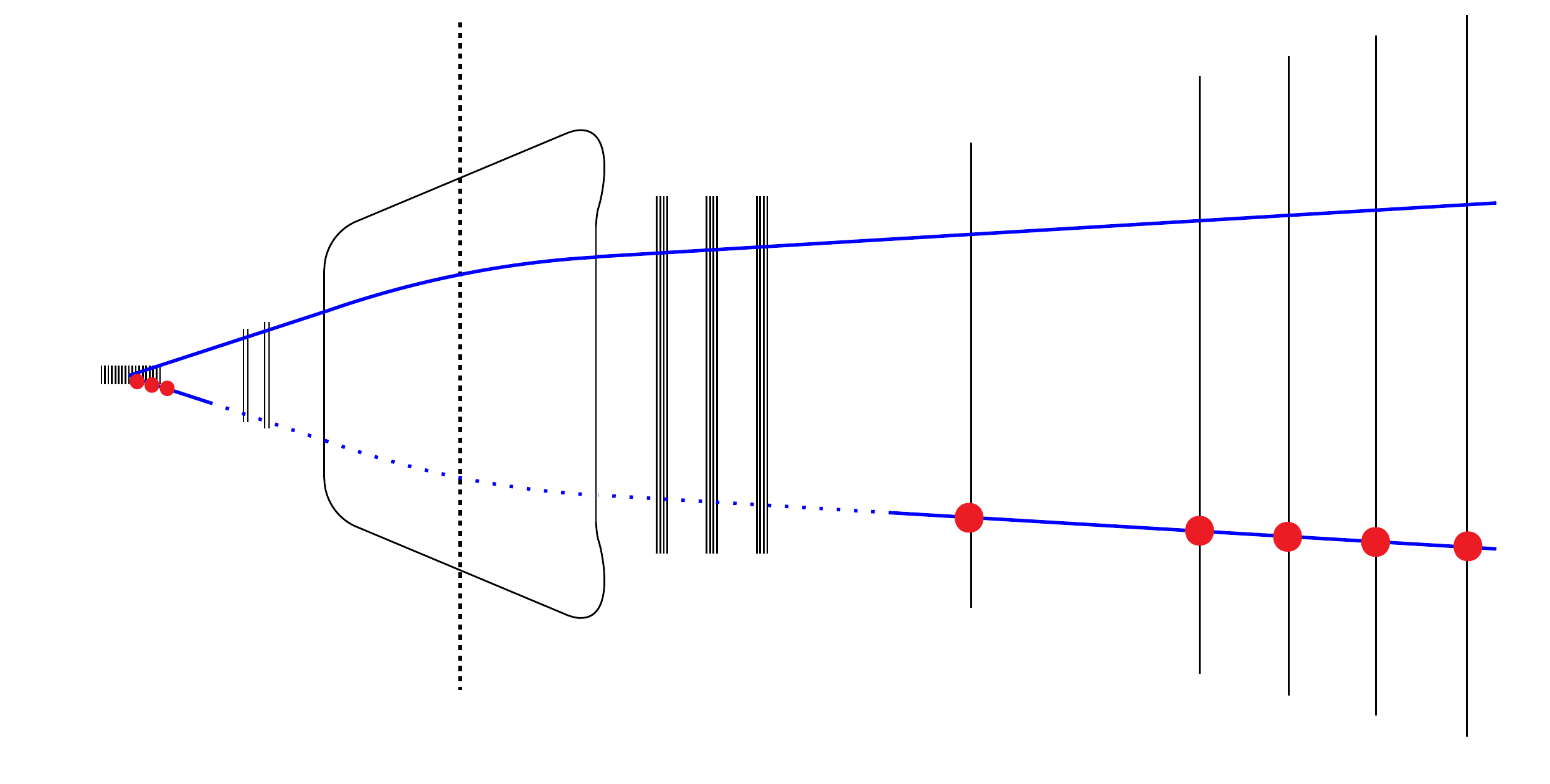}
      \put (5,40) {\small{(b)}}
    \end{overpic}
  \end{minipage}
  \begin{minipage}[b]{0.65\linewidth}
    \begin{overpic}[width=0.95\textwidth]{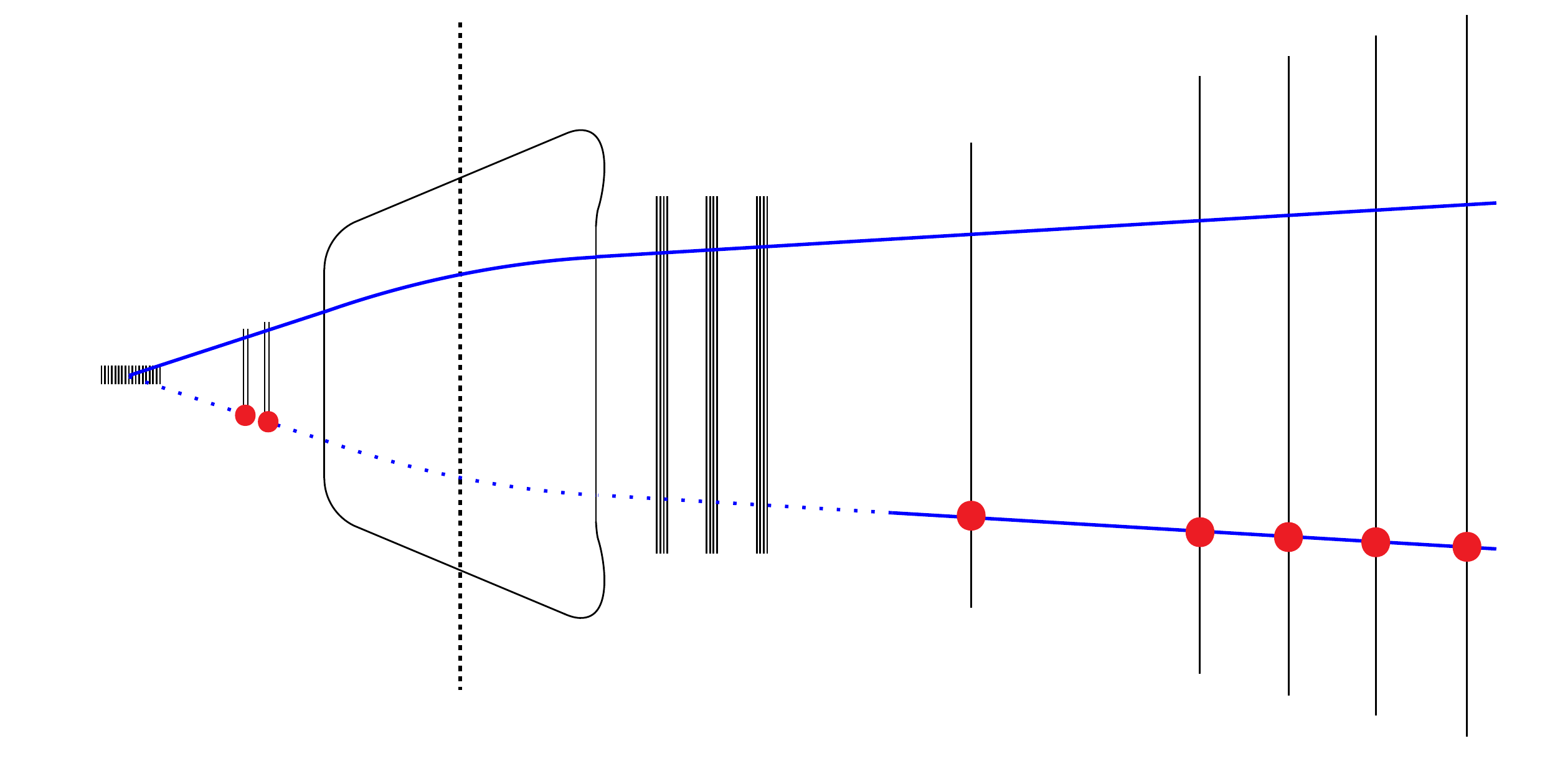}
      \put (5,40) {\small{(c)}}
    \end{overpic}
  \end{minipage}

  \caption{\small 
    Illustration of the three tag-and-probe methods: (a) the VELO method, (b)
    the T-station method, and (c) the long method. The VELO (black rectangle),
    the two TT layers (short bold lines), the magnet coil, the three T~stations
    (long bold lines), and the five muon stations (thin lines) are shown in all
    three subfigures. The upper solid blue line indicates the tag track, the
    lower line indicates the probe with red dots where hits are required and
    dashes where a detector is probed.}
  \label{fig:detectorsketch}
\end{figure}

The tag-and-probe method uses two-prong decays, where one of the decay
products, the ``tag'', is fully reconstructed as a long track, while the other particle, the
``probe'', is only partially reconstructed.
The
probe should carry enough momentum information that the invariant mass of
the parent particle can be reconstructed with a sufficiently high resolution. The invariant mass of the two-prong decay allows for a discrimination against background. The track reconstruction efficiency for long tracks is then obtained by matching the partially
reconstructed probe track to a long track. If a match is found, the probe track is defined
as efficient. The three methods described below all use $\jpsi\to\mup\mun$
decays, as the daughter particles have information in the muon system which can be exploited in the reconstruction of the probe track.
The approaches, however, use different combinations of tracking detectors for the
partial reconstruction of the probe track.

\subsection{VELO method}
\label{sec:velo}

The track reconstruction efficiency in the VELO is measured using downstream
tracks as probes, as illustrated in Fig.~\ref{fig:detectorsketch}(a). A downstream track and a long track of the same muon do not necessarily share all hits in the T~stations. Therefore, a probe track is considered to be found as a long track if there is a long track with at least 50\% common hits in the T~stations. In
simulated events the fraction of 50\% common hits is found to be an appropriate and stable matching criterion.

\subsection{T-station method}
\label{sec:tstation}

The measurement of the track reconstruction efficiency in the T~stations for
particles that have VELO and muon segments is illustrated in
Fig.~\ref{fig:detectorsketch}(b).  A dedicated algorithm reconstructs muons as straight tracks
starting from hits in the last muon station, see for example Refs.~\cite{LHCb-PUB-2009-017, Passaleva:2008a}. These are subsequently matched to VELO tracks.

A long track is considered to be matched to a probe track if two
requirements are met. Firstly, the probe track and the long track have to be
reconstructed from the same VELO seed. Secondly, at least two hits on the probe track
in the muon stations have to be compatible with the extrapolation of the
long track into the muon stations.
It is found in simulated events that requiring two common hits in the muon stations is sufficient to ensure compatible
trajectories of the long track and the VELO-muon probe track.

\subsection{Long method}
\label{sec:long}

The long method uses probe tracks that have hits in the TT and in the muon
stations as illustrated in Fig.~\ref{fig:detectorsketch}(c).
This method measures the efficiency to reconstruct long tracks because the
long-track-finding algorithms do not require the presence of TT hits. Therefore, the efficiency to find a long track is, to first order, independent of the efficiency to find such a TT-muon track.
These (TT-muon)
tracks are found by a dedicated reconstruction of tracks in the muon stations,
which are subsequently matched to TT hits. A TT-muon track
is considered to be reconstructed as a long track in case more
than 70\% of the hits in the muon stations are compatible with the extrapolation of the
long track into the muon stations. In case the long
track has TT hits, it needs to share at least 60\% of the TT hits as well. These
fractions have been optimised in simulation and the results are stable with respect to small differences in data and simulation.

\section{Trigger and selection requirements}

\begin{table}
  \begin{center}
  \caption{\small Settings of the software trigger selection as a function of
    data taking period. Only the tag muon is required to pass the selection. For more information see Refs.~\cite{LHCb-PUB-2011-006,LHCb-PUB-2011-003,LHCb-DP-2012-004,trigger2012}.}
  \label{tab:trigger}
  {\small
  \begin{tabular}{llll} \hline
    2010  & 2011 & 2012  & 2012 \\
          &      & (first $\unit{0.7}{\invfb}$) & (remaining $\unit{1.3}{\invfb}$) \\\hline
$\ip>0.11\mm$   & $\ip>0.5\mm$      & $\ip>0.5\mm$ & $\ip>0.5\mm$ \\
$\chisqip>16$ & $\chisqip>200$ & $\chisqip>200$ &$\chisqip>200$ \\
$p>8.0\gevc$      &  $p>8.0\gevc$& $p>8.0\gevc$&$p>3.0\gevc$ \\
$\pt>1.0\gevc$    & $\pt>1.3\gevc$ & $\pt>1.3\gevc$& $\pt>1.3\gevc$  \\
$\chisqndf(\text{track})<2$  & $\chisqndf(\text{track})<2$ & $\chisqndf(\text{track})<2.5$&$\chisqndf(\text{track})<2.5$  \\ \hline
  \end{tabular}
  }
  \end{center}
\end{table}

The candidate decays are first required to pass a hardware trigger, which selects
muons in the muon system with a transverse momentum, $\pt>1.48\gevc$, or
dimuons where the product of the two transverse momenta is greater than
$\pt_1\times\pt_2>(1.296\gevc)^2$. In 2012 these thresholds have been raised to
$\pt>1.76\gevc$ and $\pt_1\times\pt_2>(1.6\gevc)^2$, respectively. The reconstruction of both
muons in the hardware trigger does not bias the determination of the track reconstruction efficiency since it does not use
information from the tracking system (VELO, TT, and T~stations).

The subsequent software stage reconstructs the tag muon in the entire tracking system
and in the muon system. The tag muon is required
to have high \pt, high $p$, large \ip and \ipchisq with
respect to all PVs in the event,
where \chisqip is defined as the difference
in \chisq of a given PV reconstructed with and without the considered track. Furthermore,
a good $\chi^2$ per degree of freedom (\chisqndf) of the trigger track fit is required. Different
selection criteria are used during data taking as listed in Table~\ref{tab:trigger} to fit
different data taking conditions.
The \ip and \chisqip requirements restrict the sample to \jpsi originating from \Pb hadron decays.
Only the tag muon is required to be reconstructed in the software trigger in order to avoid any bias on the track reconstruction efficiency, caused by fully reconstructing the two-prong decay with two long tracks.

\begin{table}
  \begin{center}
  \caption{\small Selection requirements on the tag and probe tracks and on the
  combination into a $\jpsi$ candidate for the three different methods.}
  \label{tab:selection}
  {\small
  \begin{tabular}{l|l|l|l} \hline
             & {VELO}           & {T-station}        & {Long}     \\
             & {method}         & {method}           & {method}   \\\hline
Tag          & \multicolumn{3}{c}{Long track} \\
             & \multicolumn{3}{c}{used in single muon trigger} \\\cline{2-4}
             &                  & $\dllmupi>2$
                                                     & $\dllmupi>2$\\
             & $\chisqndf(\text{track})<5$ & $\chisqndf(\text{track})<3$   & $\chisqndf(\text{track})<2$       \\
             & $p>5.0\gevc$     & $p>7.0\gevc$       & $p>10\gevc$            \\
             & $\pt>0.7\gevc$   & $\pt>0.5\gevc$     & $\pt>1.3\gevc$ \\
             &                  &                    & $\ip>0.5\mm$    \\ \hline
Probe        & Downstream track & VELO-muon track    & TT-muon track  \\
             & $p>5.0\gevc$     & $p>5.0\gevc$       & $p>5.0\gevc$   \\
             & $\pt>0.7\gevc$   & $\pt>0.5\gevc$     & $\pt>0.1\gevc$ \\ \hline
$\jpsi$      & $M_{\mu\mu}\in[2.9,3.3]\gevcc$ 
                                & $M_{\mu\mu}\in[2.7,3.5]\gevcc$ 
                                                    & $M_{\mu\mu}\in[2.6,3.6]\gevcc$ \\
                                                    & $\chisqndf(\text{vertex})<5$ & $\chisqndf(\text{vertex})<5$ & $\chisqndf(\text{vertex})<5$         \\
             & $N_\jpsi=1$      & $N_\jpsi=1$      & $N_\jpsi=1$             \\
             &                  & $p>7.0\gevc$     & $\ip<0.8\mm$             \\
            \hline
  \end{tabular}
  }
  \end{center}
\end{table}

Further selection criteria are applied as listed in
Table~\ref{tab:selection}: the \chisqndf from the
track fit of the tag tracks must be small to reduce the number of fake tracks. Tag tracks have to fulfil the standard
muon selection, which requires hits in the muon stations in a search window
around the track extrapolation as explained in Ref.~\cite{LHCb-DP-2013-001}.
Both the tag and probe tracks have minimal $p$ and \pt requirements to remove badly reconstructed tracks and combinatorial background.
 In order to remove contamination from hadrons, the particle identification system is used.
The differences between the logarithm of the likelihood of the tag to be a muon and to be a pion, \dllmupi, is computed and only tag tracks with a high \dllmupi are used. The range of the invariant mass of the
$\mup\mun$ combination, $M_{\mu\mu}$, is chosen sufficiently large to estimate
the background contribution from the mass sidebands. Finally, the $\chisqndf$
from the vertex fit of the tag- and the probe-track has to be small, in order to remove combinatorial background; 
and the number of \jpsi decays per event ($N_\jpsi$) must be one, to simplify the association procedure described in the preceding subsections.
Additionally, the T-station method only considers \jpsi candidates with a momentum greater than 7\gevc, and the long method only \jpsi candidates with an \ip smaller than 0.8\mm, as both selections are effective in reducing background contamination without biasing the efficiency determination.
After the full selection chain the sample amounts to about $6\,000$ decays for 2010 for
the long and the T method, while for the VELO method $12\,000$ decays are
selected. The 2011 and 2012 data samples comprise more than $300\,000$ decays in
total for all methods and data taking periods. 

\subsection{Mass resolution}

\begin{figure}
  \centering
  \includegraphics[width=0.46\textwidth]{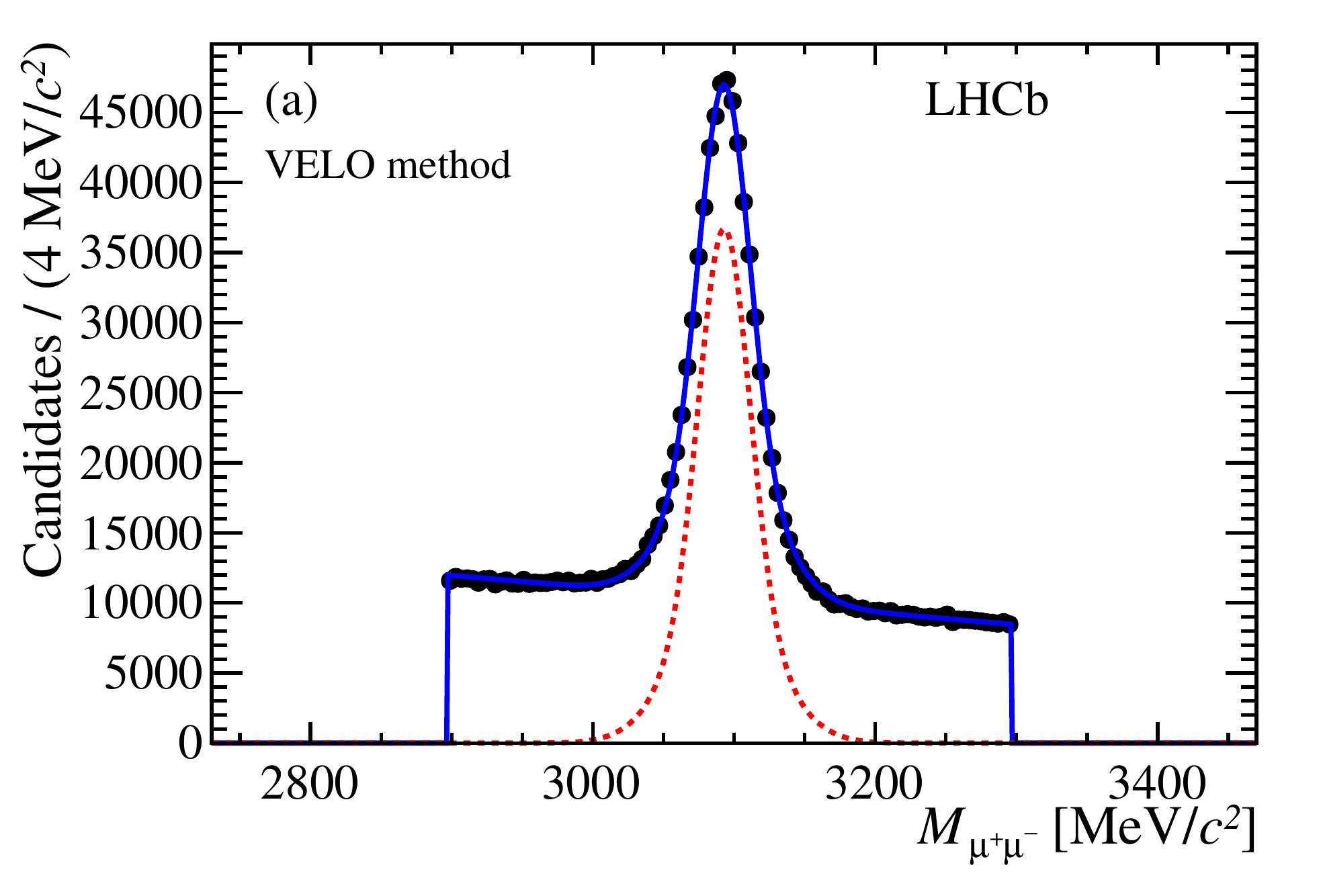}
  \includegraphics[width=0.46\textwidth]{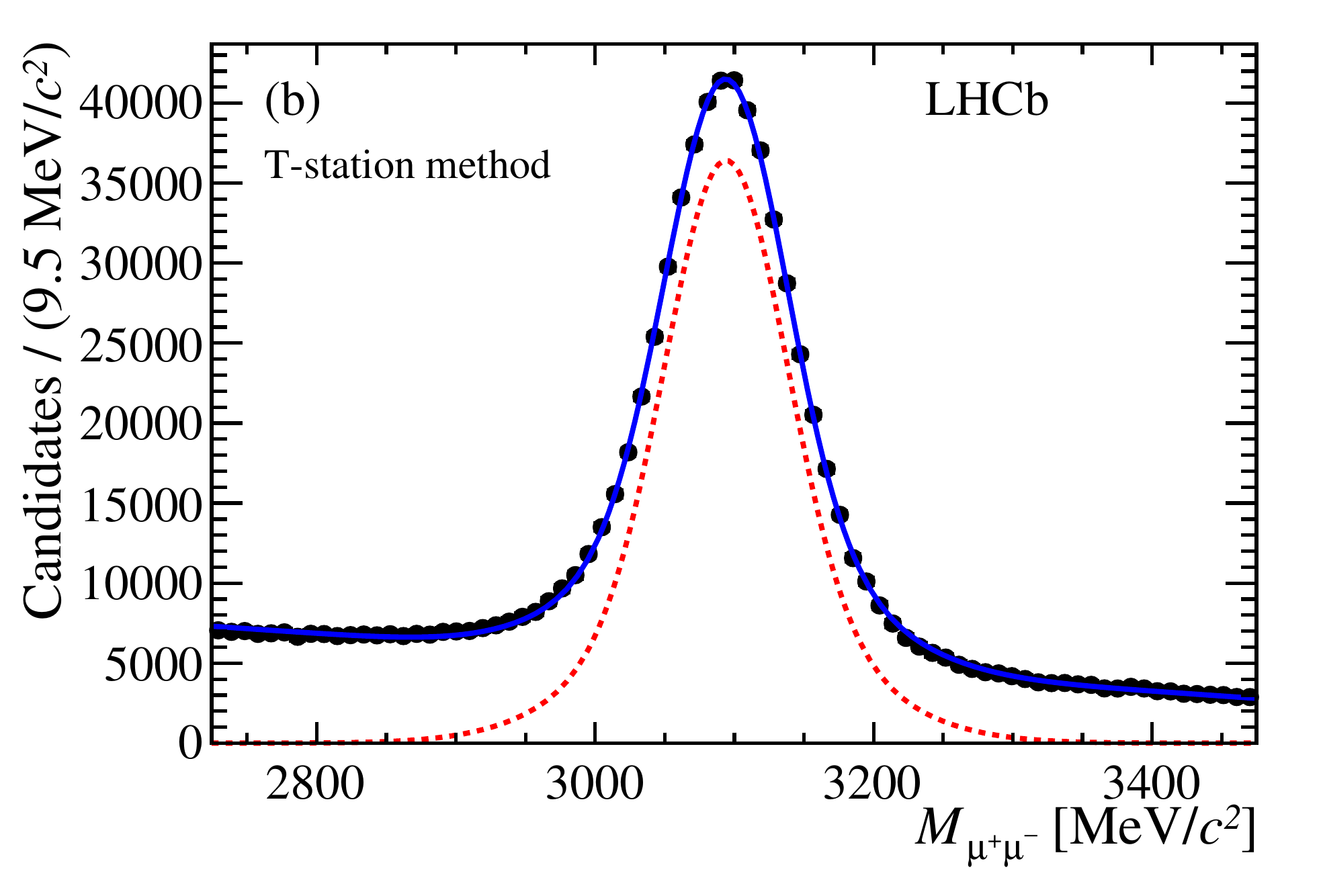}
  \includegraphics[width=0.46\textwidth]{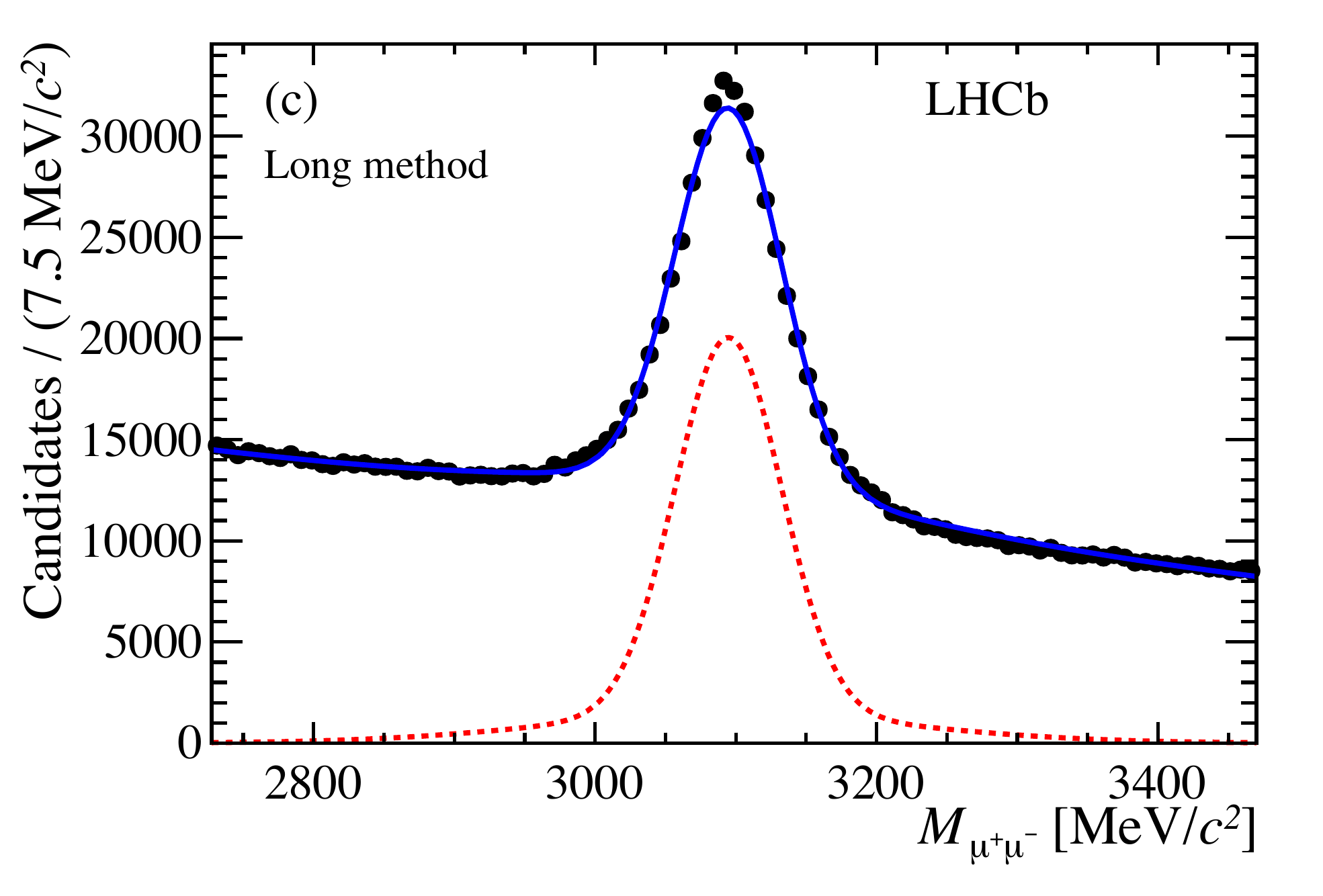}
  \includegraphics[width=0.46\textwidth]{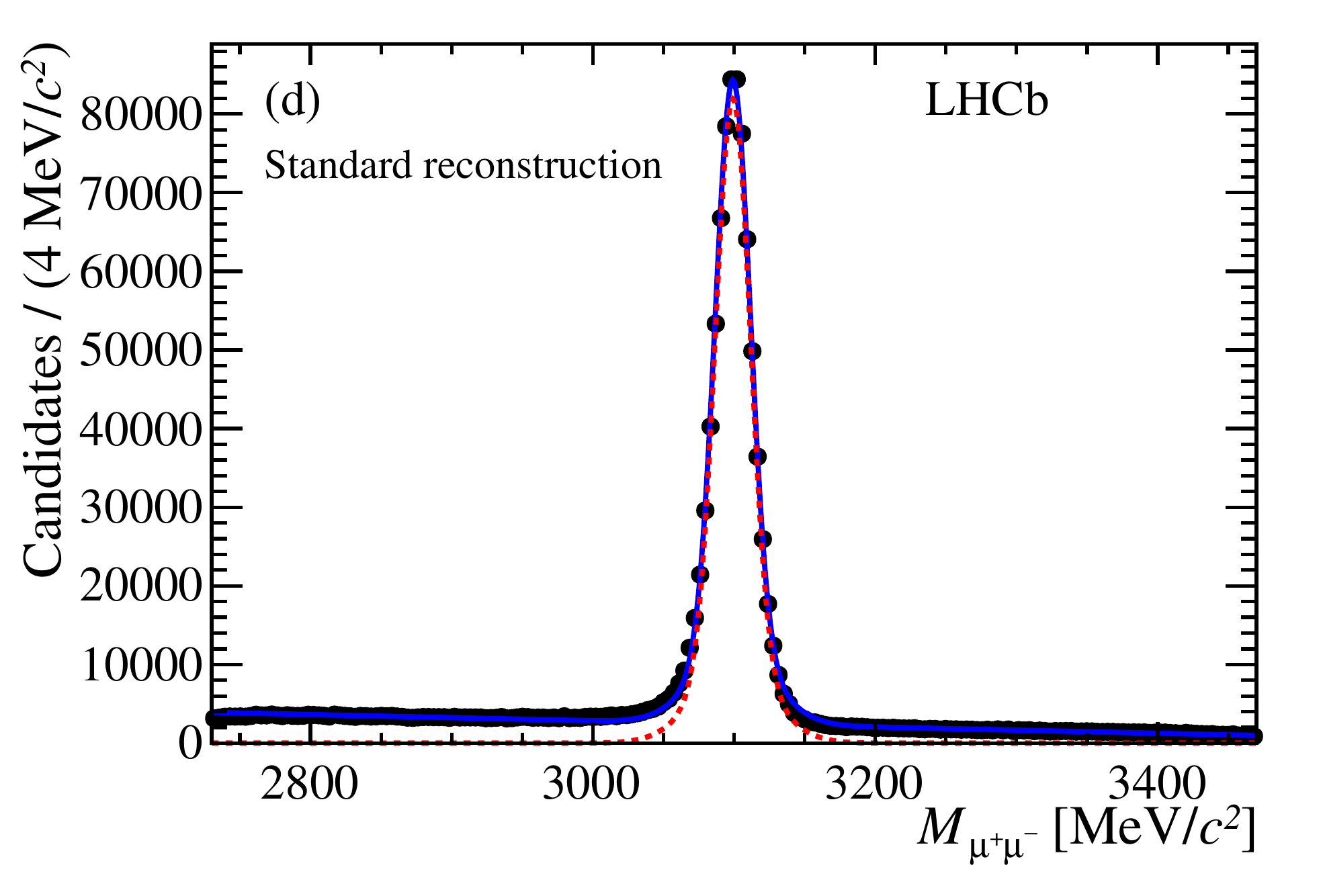}
  \caption{\small Invariant mass distributions for reconstructed \jpsi candidates from the 2011 dataset. The solid line shows the fitted distribution for signal and background, the dotted line is the signal component. The subfigures are (a) the VELO method, (b) the T-station method, (c) the long method. For comparison of resolution and signal purity (d) shows the invariant mass distribution of \jpsi candidates obtained with the standard reconstruction at LHCb.	}
  \label{fig:mass}
\end{figure}

To illustrate the mass resolutions that can be achieved, the dimuon invariant mass distributions from \jpsi candidates in the three methods are shown in Fig.~\ref{fig:mass} using the 2011 data sample.
The difference in the visible ranges in
Fig.~\ref{fig:mass}(a) compared with the other distributions in Fig.~\ref{fig:mass} is a consequence of the different dimuon invariant mass cuts as listed in Table~\ref{tab:selection}.  The invariant mass
distribution using two long tracks is shown in
Fig.~\ref{fig:mass}(d) for comparison. The signal peak is fitted with the sum of two Gaussian functions for this illustration. The effective mass resolution is about 24\mevcc for the VELO method, 57\mevcc for the T-station method and for the long method. This is to be compared to the standard reconstruction with two long tracks that achieves a resolution of 16\mevcc.

\section{Calculation of efficiency}

The track reconstruction efficiency is calculated as the fraction of
reconstructed \jpsi decays where the probe track can be matched to a long
track. To estimate the number of \jpsi decays, an unbinned extended maximum likelihood fit is performed to the mass
distributions. For the VELO and T-station methods the mass distributions are
described by a single Gaussian function for the signal and an exponential
function for the combinatorial background. This model is preferred over the aforementioned sum of two Gaussian functions to improve the fit stability when measuring the dependence of the track reconstruction efficiency on kinematic variables and other event parameters. For the long method, a Crystal Ball
function~\cite{Skwarnicki:1986xj} is used for the signal, to take the tail on
the left-hand side of the mass peak into account. Since the number of decays in the
2010 data is relatively low, in this case a simple sideband subtraction is applied for the
VELO and T-station methods. All shape parameters were allowed to vary in the fit for the denominator of the efficiency; they were constrained to the found values for the numerator of the efficiency. This procedure was performed to stabilise the fit, as no difference in the shape of the numerator and denominator could be observed.
It has been checked that the choice of the model for the mass distribution has a negligible effect on the efficiency determination.

\label{sec:combinedtagandprobe}

The efficiencies obtained from the VELO and T-station methods are assumed to be
uncorrelated, aside from effects due to dependencies on the track kinematics and
the event multiplicity. The data sample is binned in kinematic variables and $N_{\rm track}$ to combine the VELO and T-station efficiencies. The efficiencies obtained with the VELO and T-station methods can be
multiplied in each bin to obtain the efficiency for finding long tracks. This
combined efficiency can be compared with the efficiency found by the long
method, giving two independent methods to probe the long track reconstruction
efficiency. 

There are, however, small differences between these two
approaches. The long method measures the efficiency for tracks that pass through TT. In the combined method, only the VELO method requires this.
Furthermore, both the VELO method and T-station method include the efficiency
that, given that both the VELO and the T-station segment tracks are
reconstructed, the corresponding long track is found.  Therefore, in the
combined efficiency, this so-called matching efficiency is counted twice. All
these effects can lead to small differences in the measured long-track
efficiency. For this reason, the ratio between the efficiencies in data and
simulation is used to compare the methods, as these uncertainties are common
for simulated and real decays and cancel when the ratio of efficiencies is
formed.

On simulated events the track reconstruction efficiency is commonly defined as
the fraction of simulated charged particles with sufficient hits in the VELO and T
stations that can be associated to a track that shares at least $70\%$ of the
hits in each participating subdetector with this particle. For all methods, this so-called hit-based efficiency in simulation agrees
within 1\% with the efficiency measured with the tag-and-probe methods. Furthermore the matching efficiency was determined to be very close to 100\%. The very small matching inefficiency does not affect the agreement between the hit-based efficiency and the tag-and-probe based efficiency in simulation. By taking the ratio between the efficiencies on data and simulation, these
discrepancies are reduced to a negligible level.

\section{Efficiency dependencies}
\label{sec:tagandproberesults}

Using the momentum spectrum of the \jpsi decay products obtained with the VELO method from data as a benchmark,
the average track reconstruction efficiency for long tracks is measured
to be $(95.4\pm0.7)\%$ for 2010 data, $(97.78\pm0.07)\%$ for 2011 data and
$(96.99\pm0.05)\%$ for 2012 data. All results confirm the good performance of the \lhcb tracking system. 
The uncertainties on these numbers are statistical only; they are binomial errors with additional terms to account for the statistical uncertainty on the number of background events. 
Systematic uncertainties are discussed in Sect.~\ref{sect:sysuncertainty}. The difference in the efficiencies between the three years is a consequence of changes in the track reconstruction and the higher centre-of-mass energy, leading to a higher track multiplicity and hence lower reconstruction efficiency for the 2012 running period. Dependencies on
the polarity of the dipole magnet, the charge of the muons, and kinematic
properties as well as the agreement with the simulation are investigated in
further detail in the following subsections.

\subsection{Comparison of magnetic field polarities}

The track reconstruction efficiencies determined from the long method are split up
into positively and negatively charged muons and into the two different magnetic field
polarities (named up and down). The results are summarised in Table~\ref{tab:magnet}.
They show compatible numbers for magnetic field up and down and for positive and
negative muons. 

For data from 2011 and 2012 there is no difference between positive and
negative muons or between the different magnet polarities. In 2010 data, a
$2.3\,\sigma$ difference between the different magnet polarities is observed
for positive muons.
No unambiguous source
of the difference is found.

\begin{table}
  \begin{center}
  \caption{\small Track reconstruction efficiencies in \% for the individual running periods using the
    long method for positive and negative muons and different magnetic field
    polarities (statistical uncertainties only).}
  \label{tab:magnet}
  {\small
  \begin{tabular}{l|c|c|c|c} \hline
 & \multicolumn{2}{c|}{Magnet up} & \multicolumn{2}{c}{Magnet down}\\
\cline{2-5}
Data & {Positive} & {Negative}  & {Positive} & {Negative} \\ \hline
2010      & $94.1\pm1.3$ & $96.0\pm1.3$ 
          & $99.3^{+0.7}_{-1.8}$ & $98.4^{+1.6}_{-1.7}$ \\
2011      & $97.0\pm0.3$ & $97.3\pm0.3$ 
          & $97.2\pm0.3$ & $97.4\pm0.3$ \\
2012      & $96.2\pm0.2$   &  $96.2\pm0.2$
          & $96.2\pm0.2$   &  $96.3\pm0.2$   \\
\hline
  \end{tabular}
  }
  \end{center}
\end{table}

\subsection{Dependencies of track reconstruction efficiency}

\begin{figure}
  \begin{center}
    \includegraphics[width=0.46\textwidth]{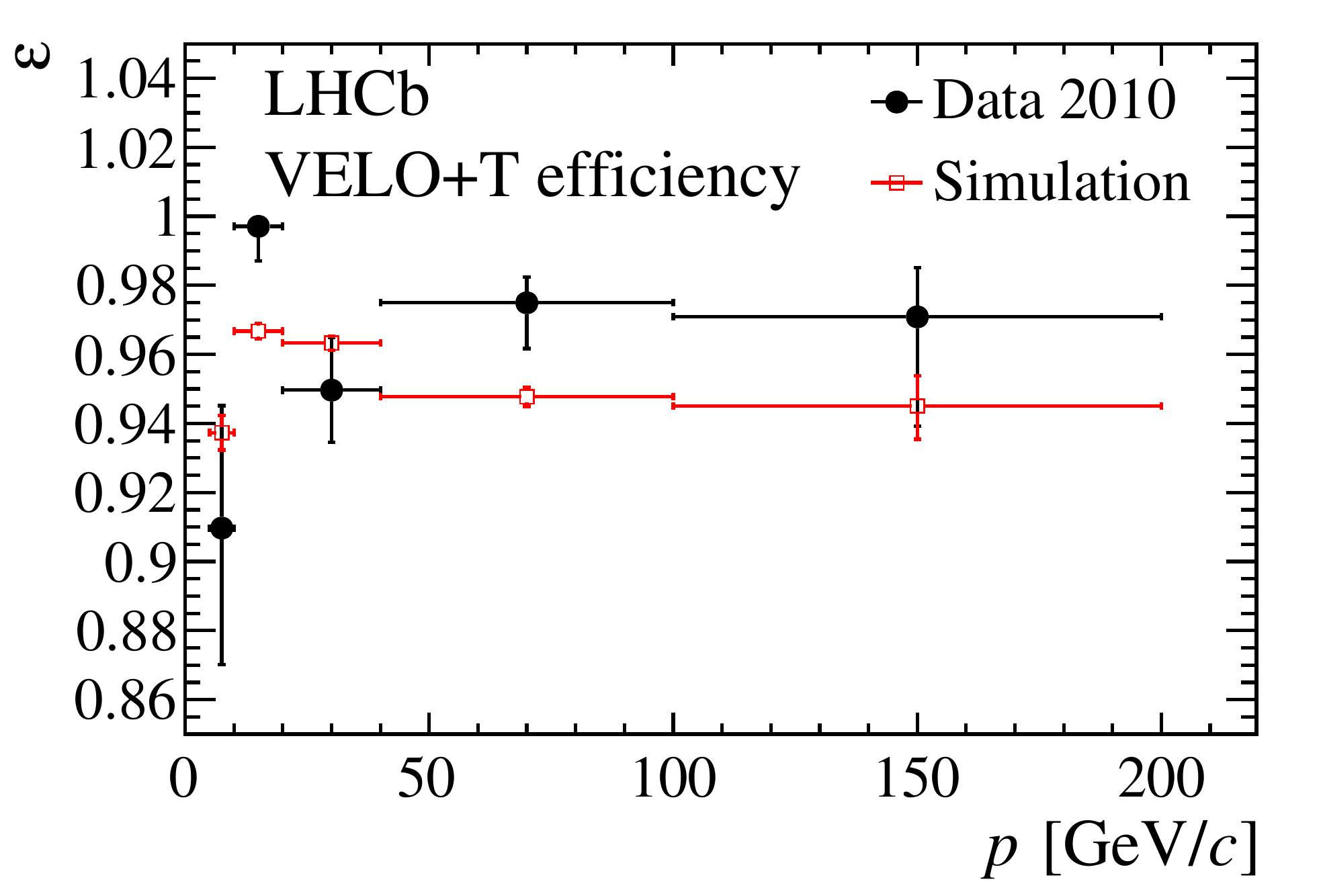}
    \includegraphics[width=0.46\textwidth]{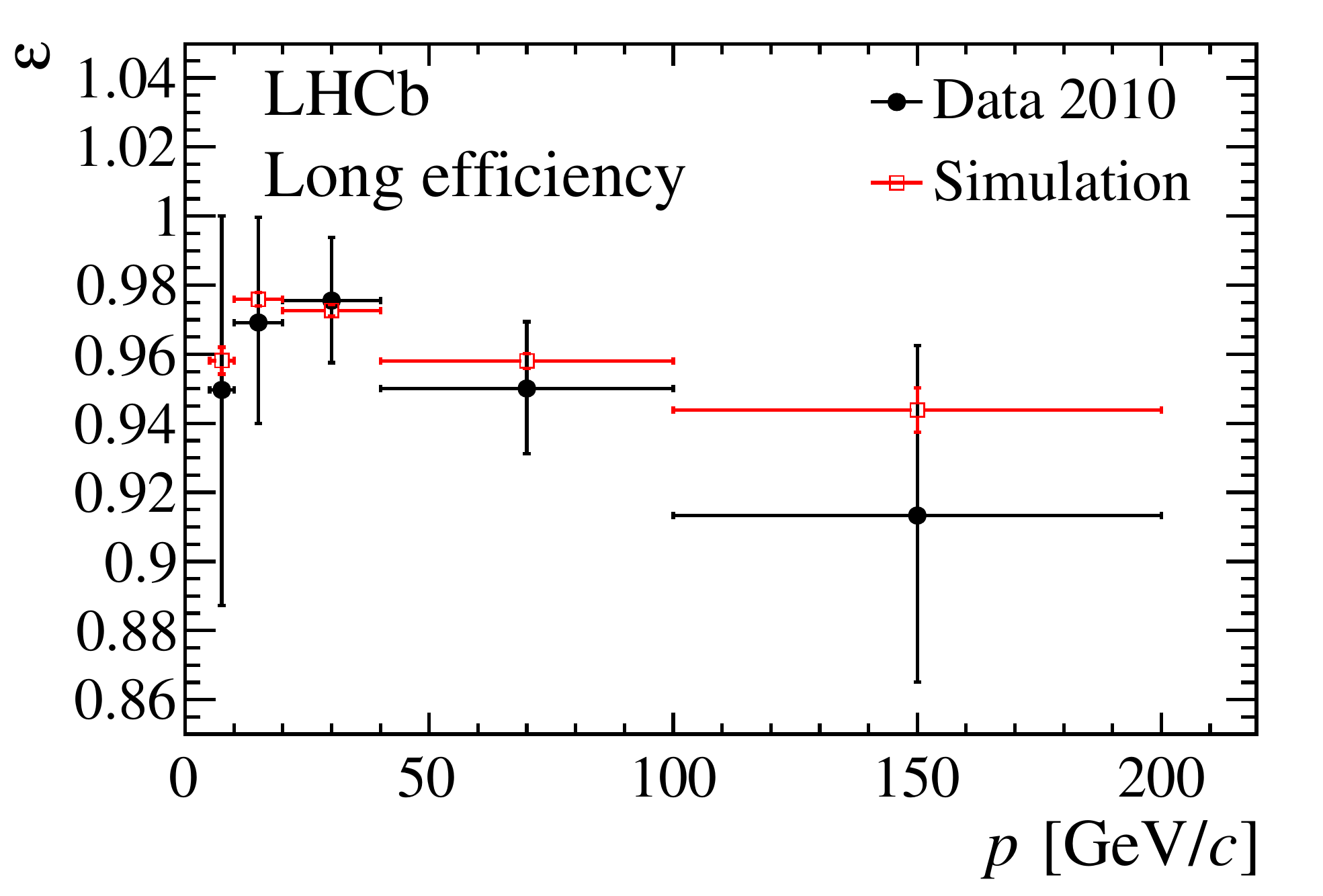}
    \includegraphics[width=0.46\textwidth]{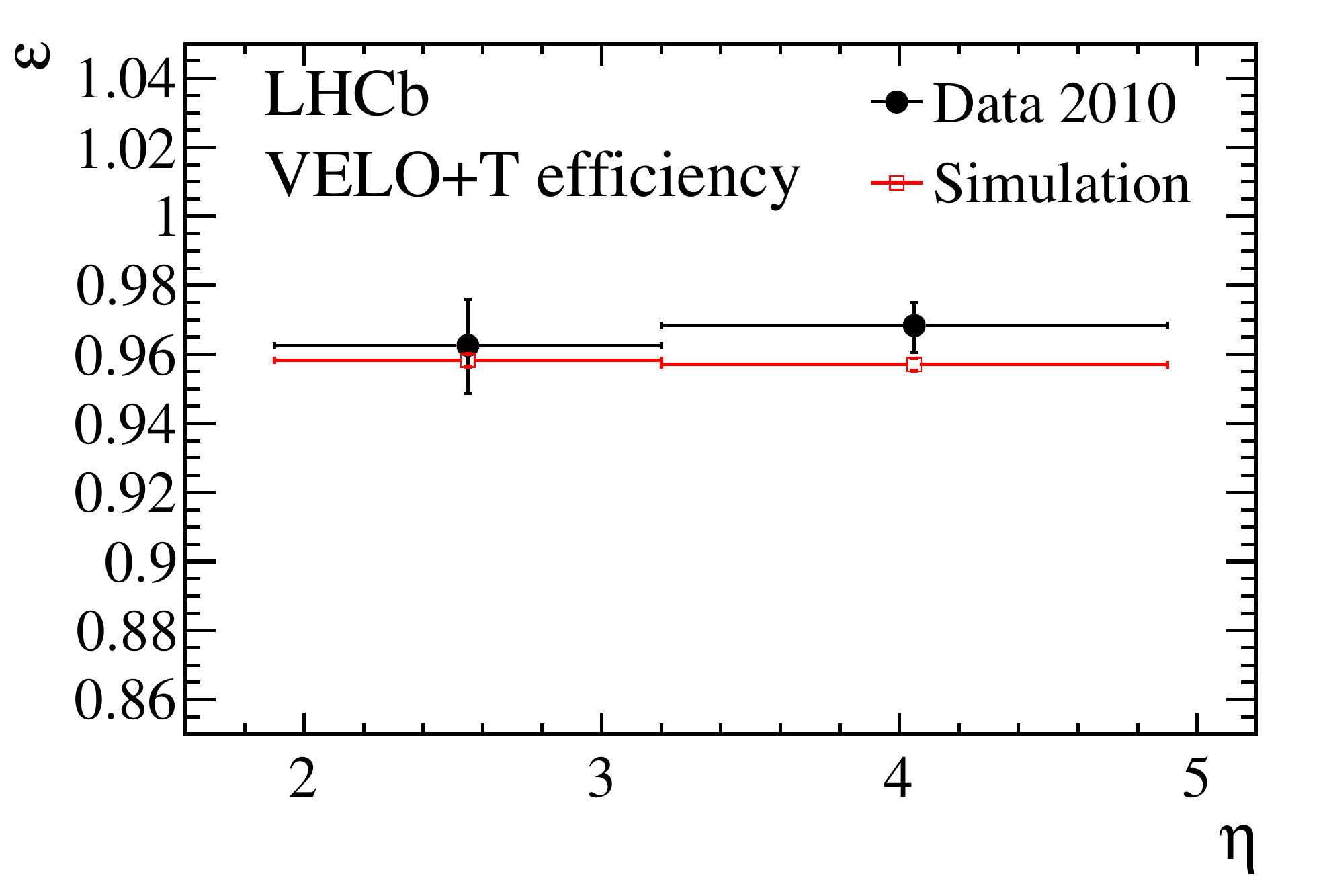}
    \includegraphics[width=0.46\textwidth]{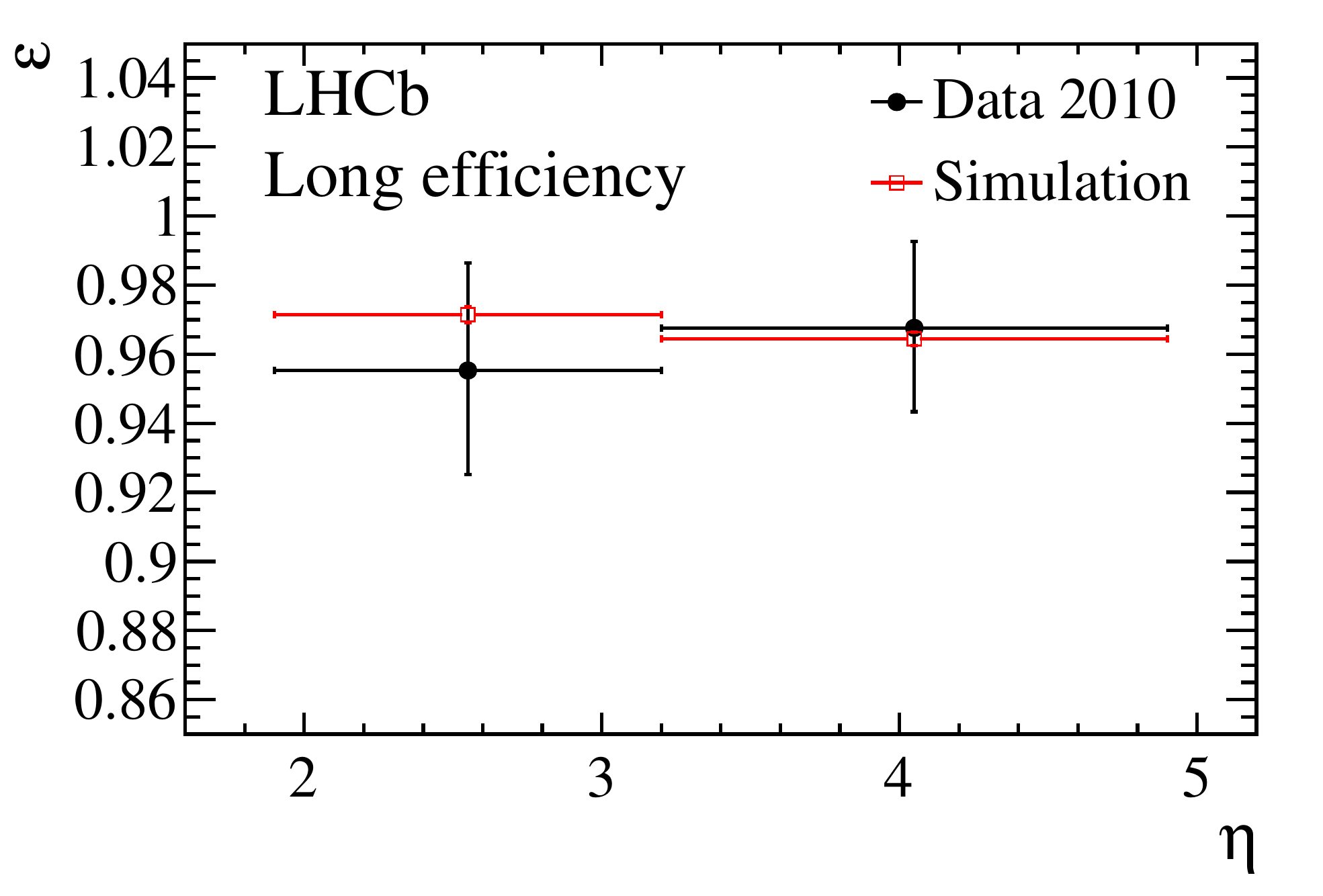}
    \includegraphics[width=0.46\textwidth]{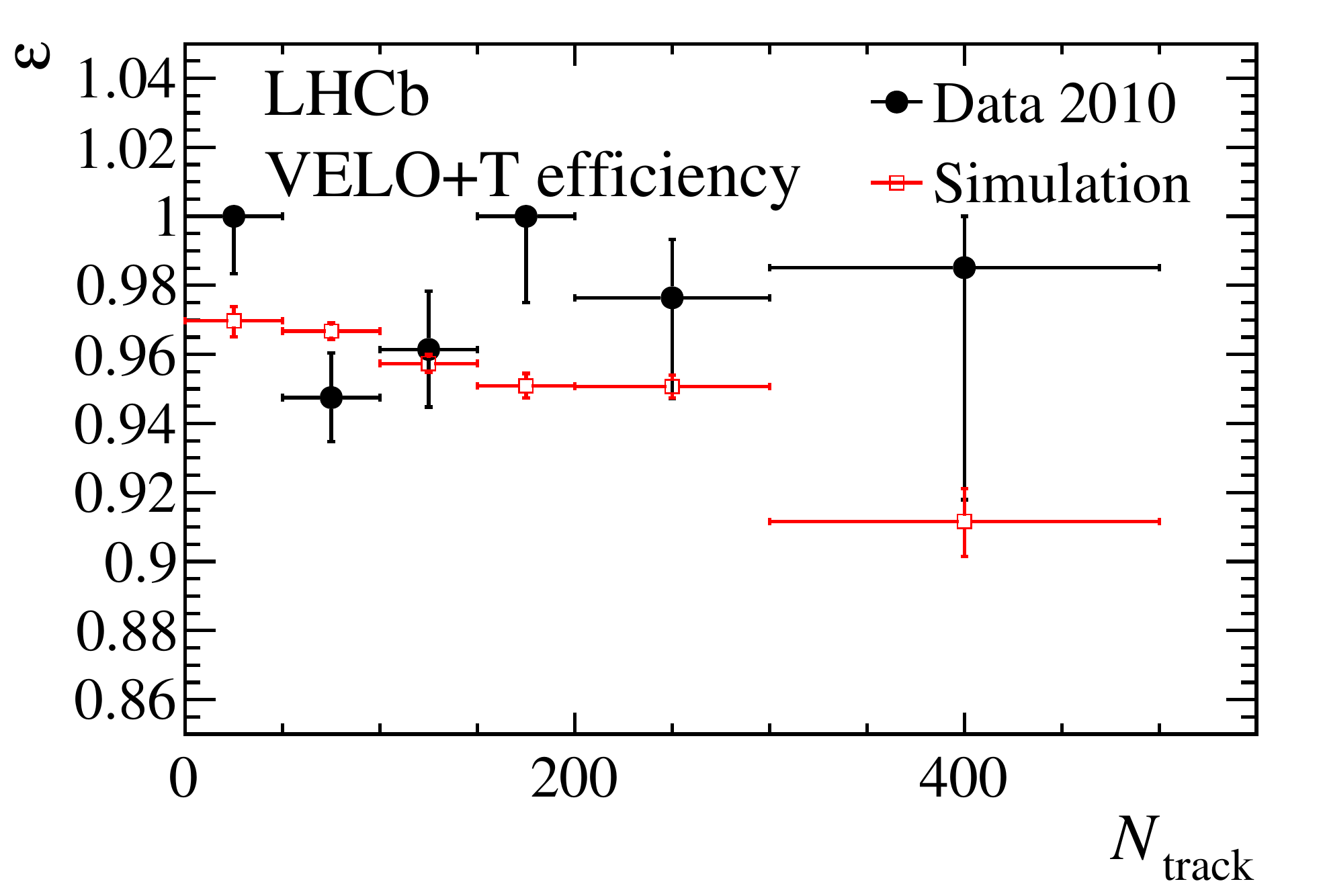}
    \includegraphics[width=0.46\textwidth]{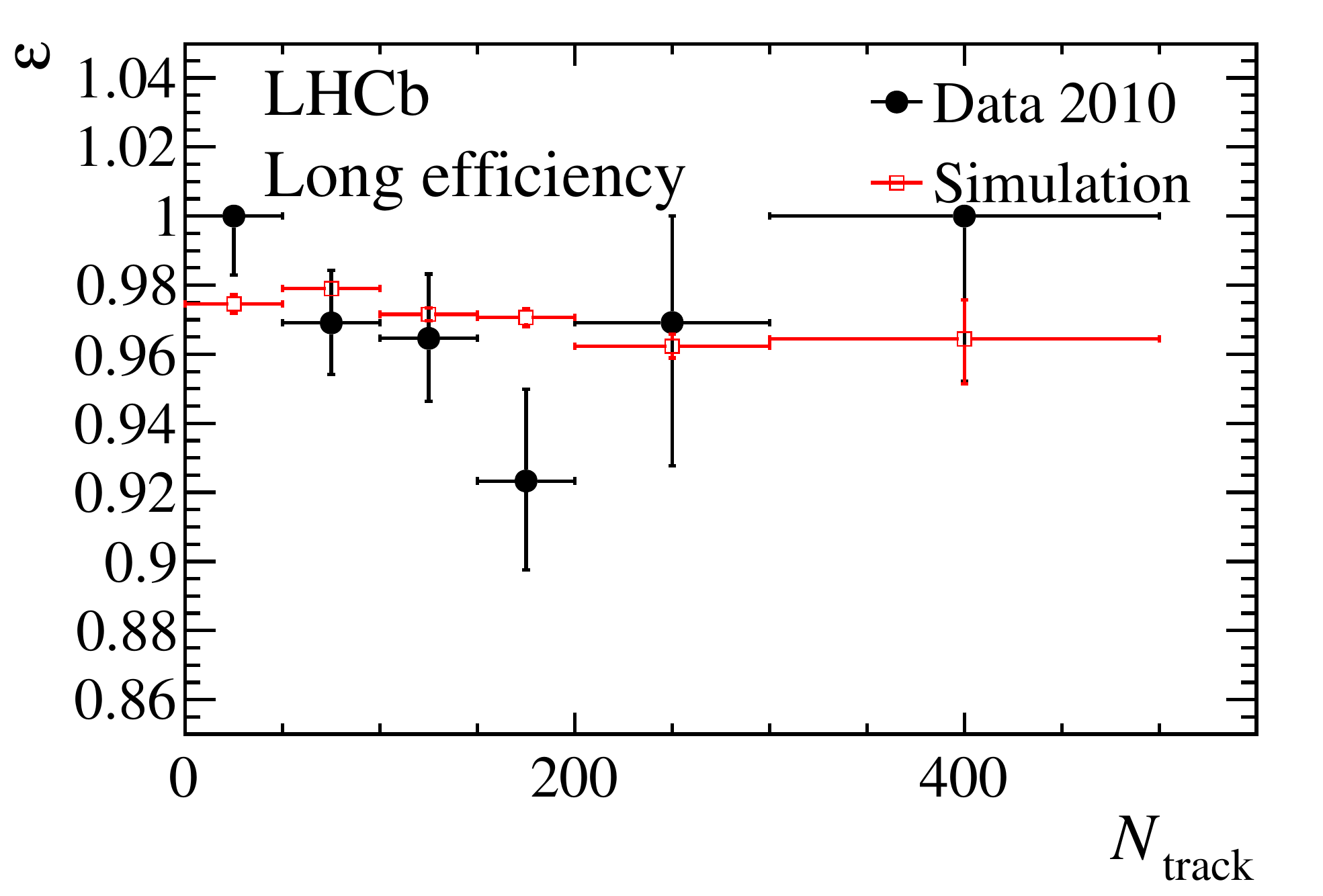}
    \includegraphics[width=0.46\textwidth]{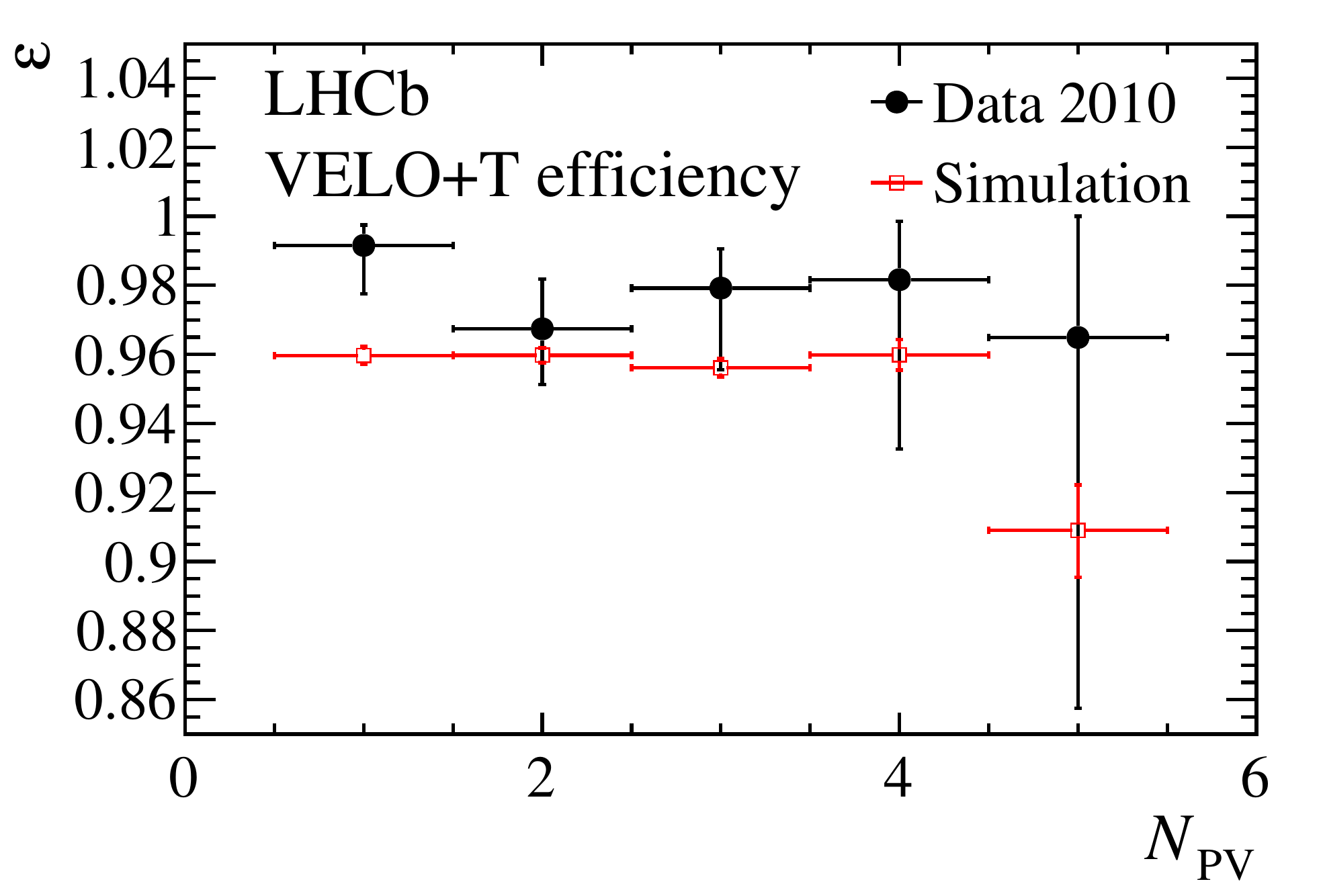}
    \includegraphics[width=0.46\textwidth]{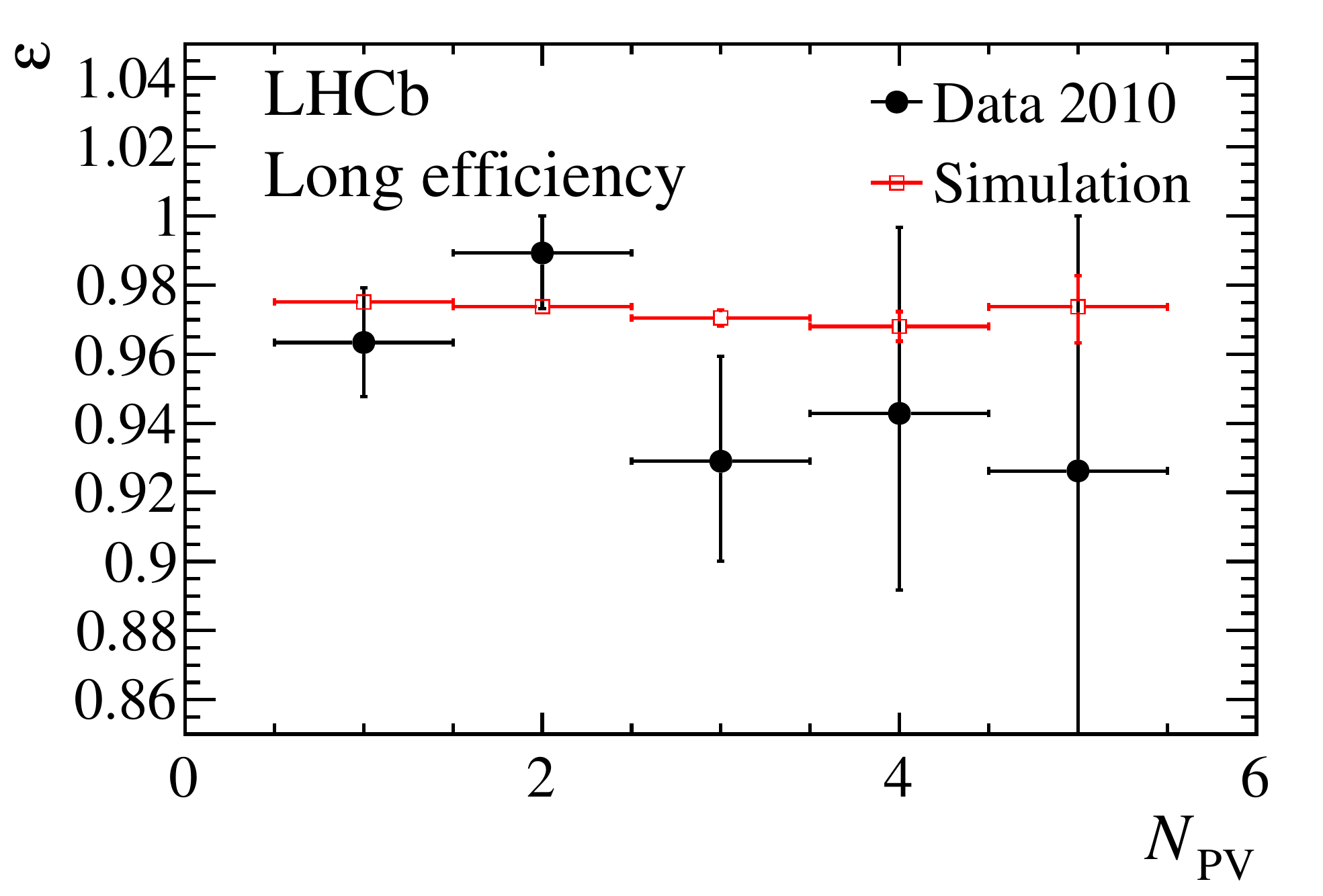}
  \end{center}
  \caption{\small Track reconstruction efficiencies for the 2010 data and for
    weighted simulation. The left-hand column shows the results of the combined
    method while the right-hand column shows the results of the long method. The
    efficiency is shown as a function of $p$ (first row), 
    $\eta$ (second row), $N_{\rm track}$ (third row), and $N_{\rm PV}$ (fourth row). The error bars indicate the
  statistical uncertainties.}
  \label{fig:effLong2010}
\end{figure}

\begin{figure}
  \begin{center}
    \includegraphics[width=0.46\textwidth]{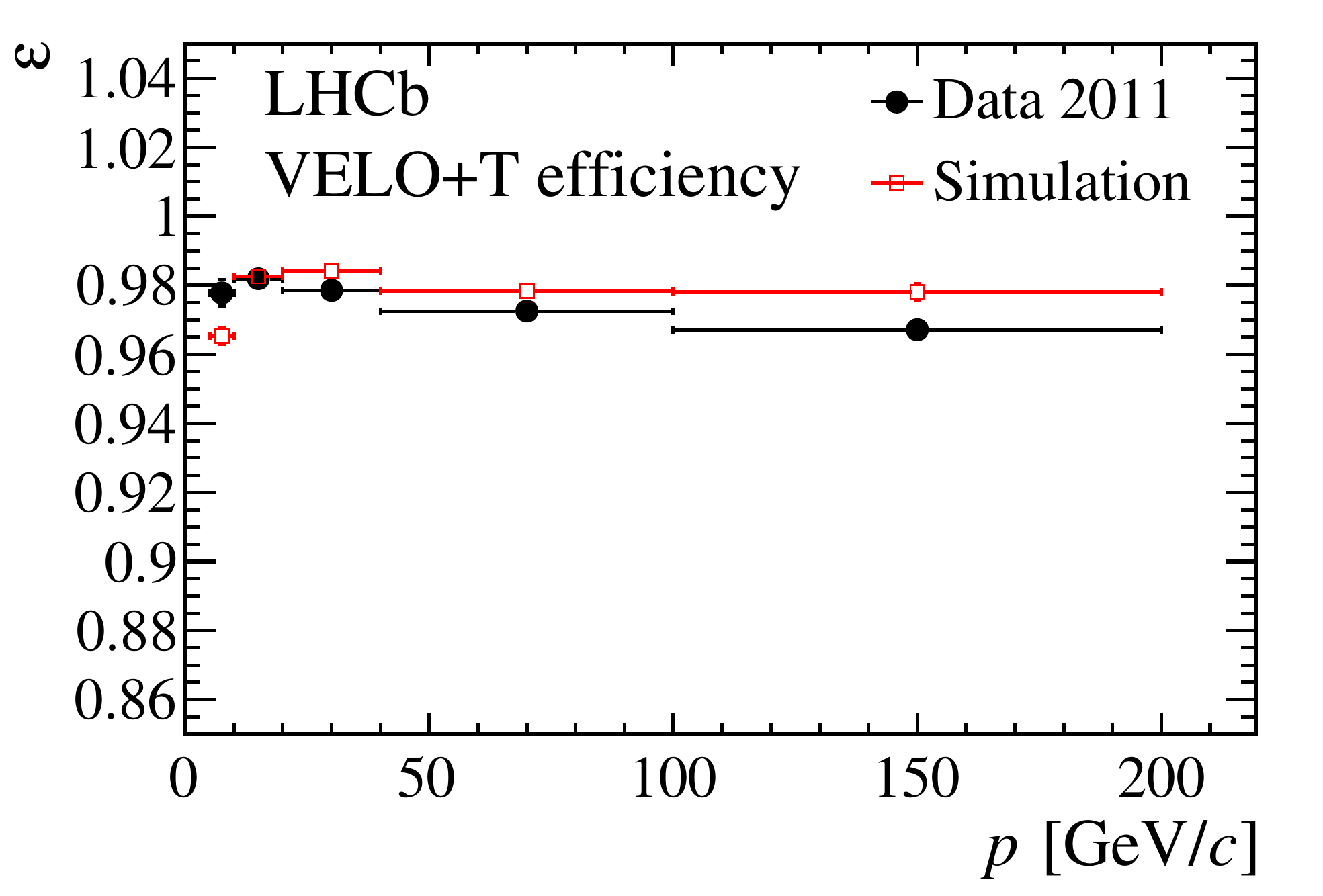}
    \includegraphics[width=0.46\textwidth]{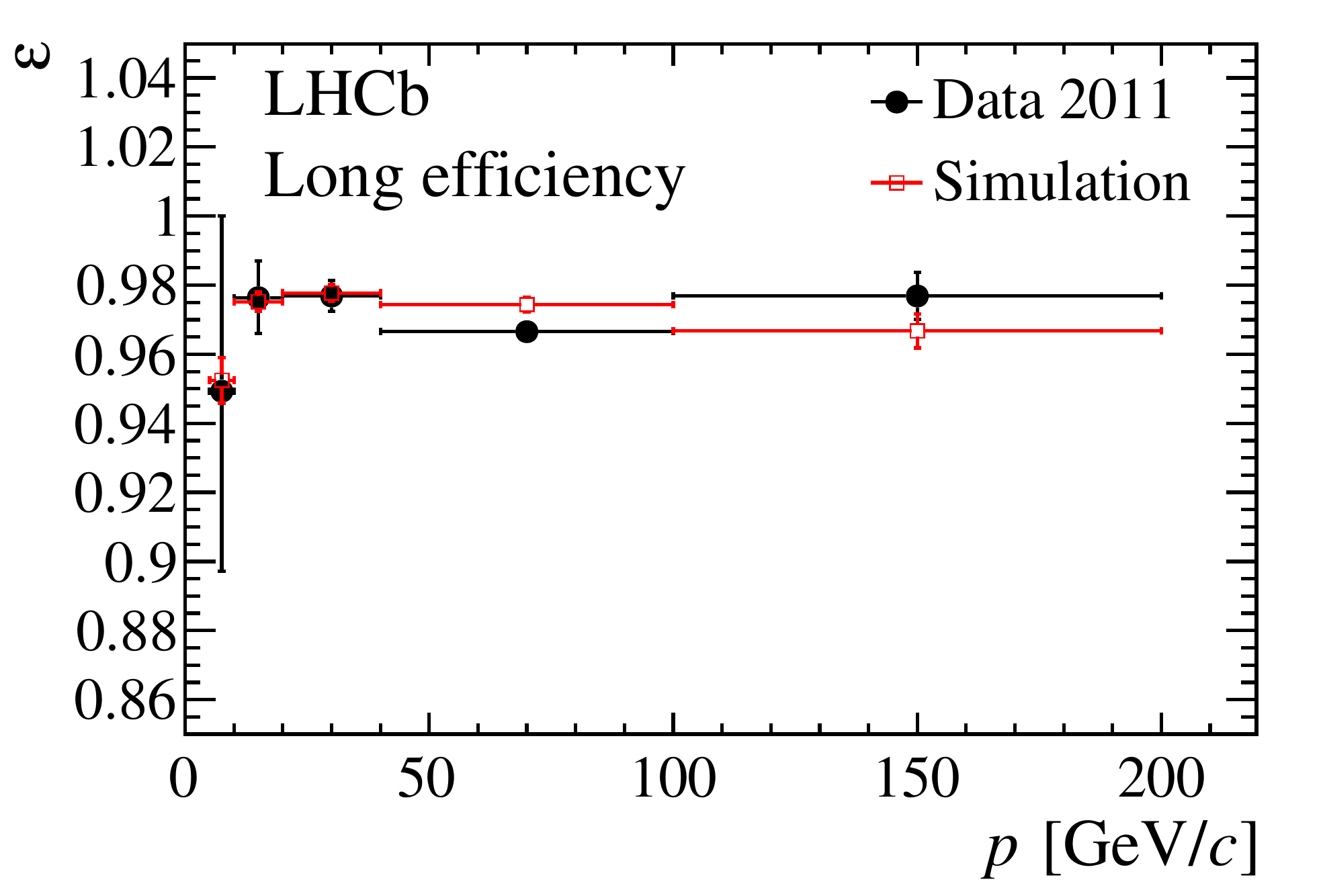}
    \includegraphics[width=0.46\textwidth]{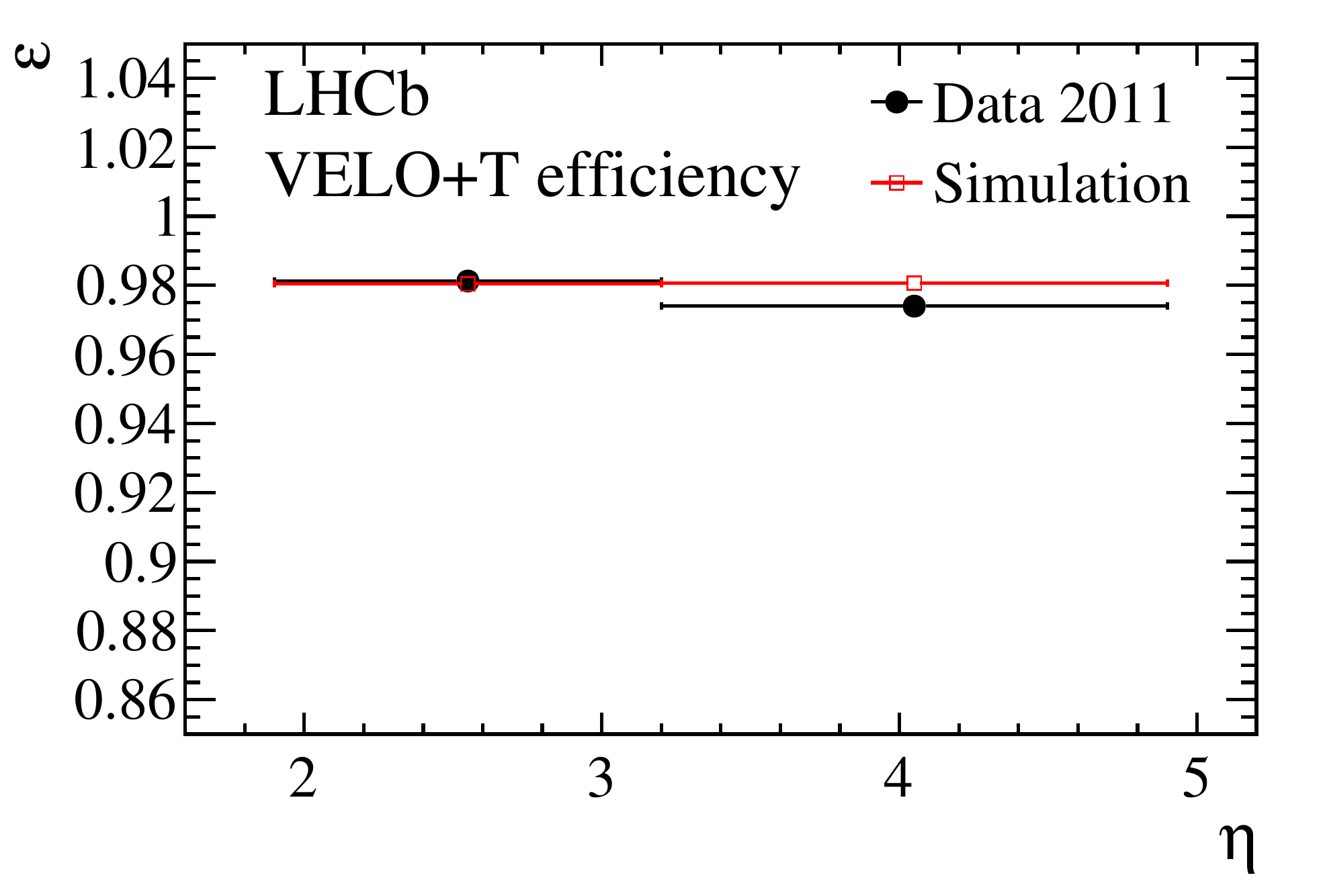}
    \includegraphics[width=0.46\textwidth]{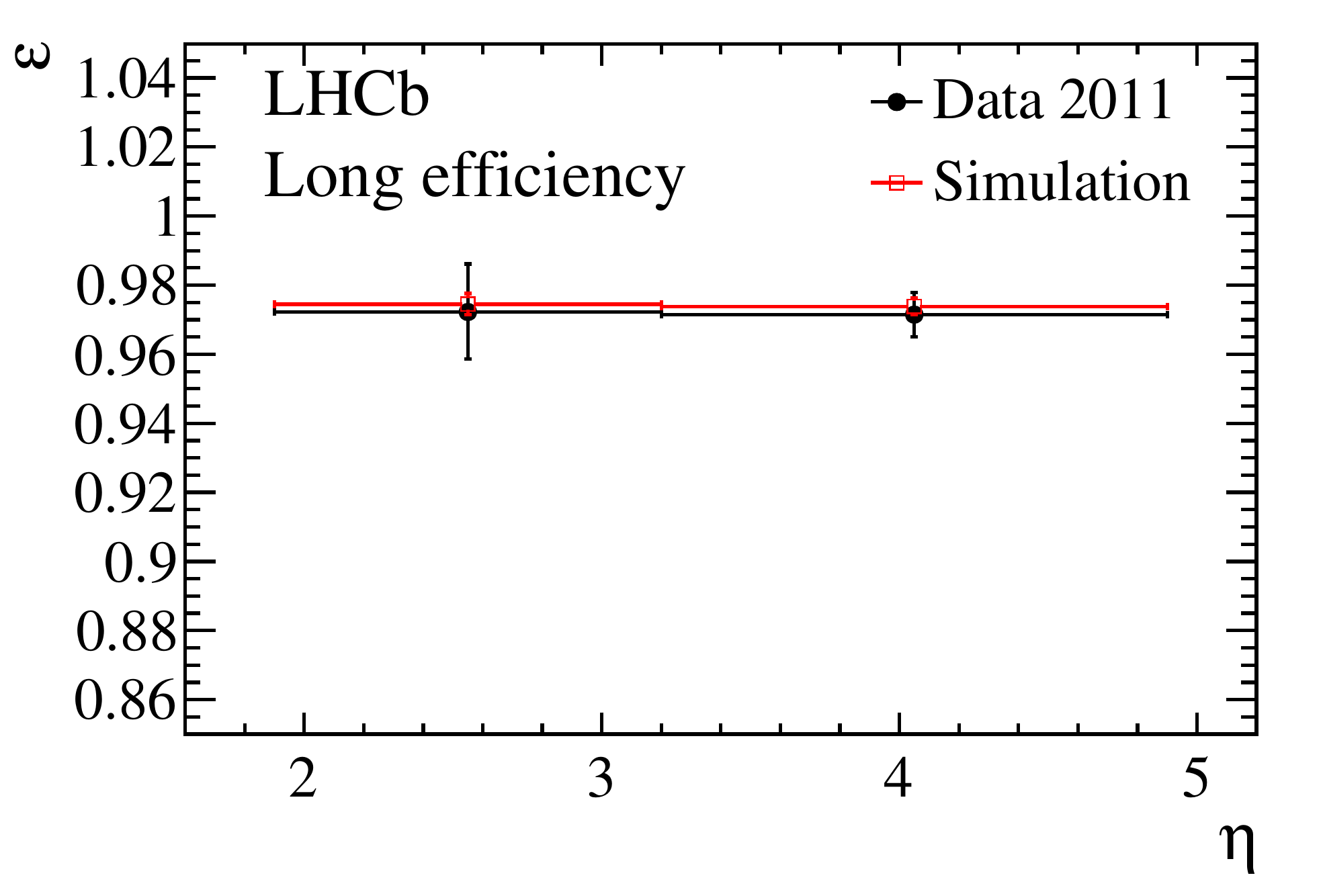}
    \includegraphics[width=0.46\textwidth]{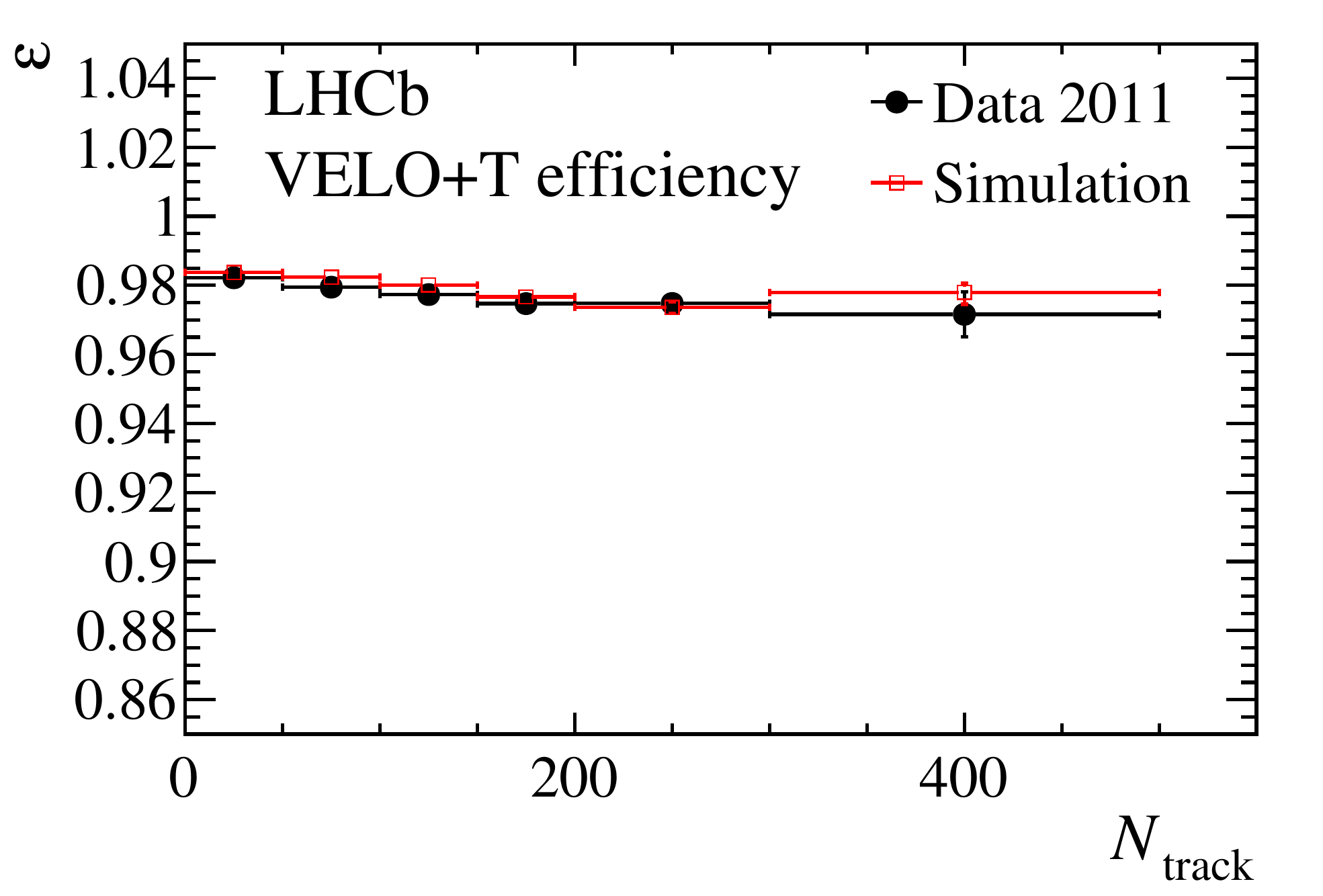}
    \includegraphics[width=0.46\textwidth]{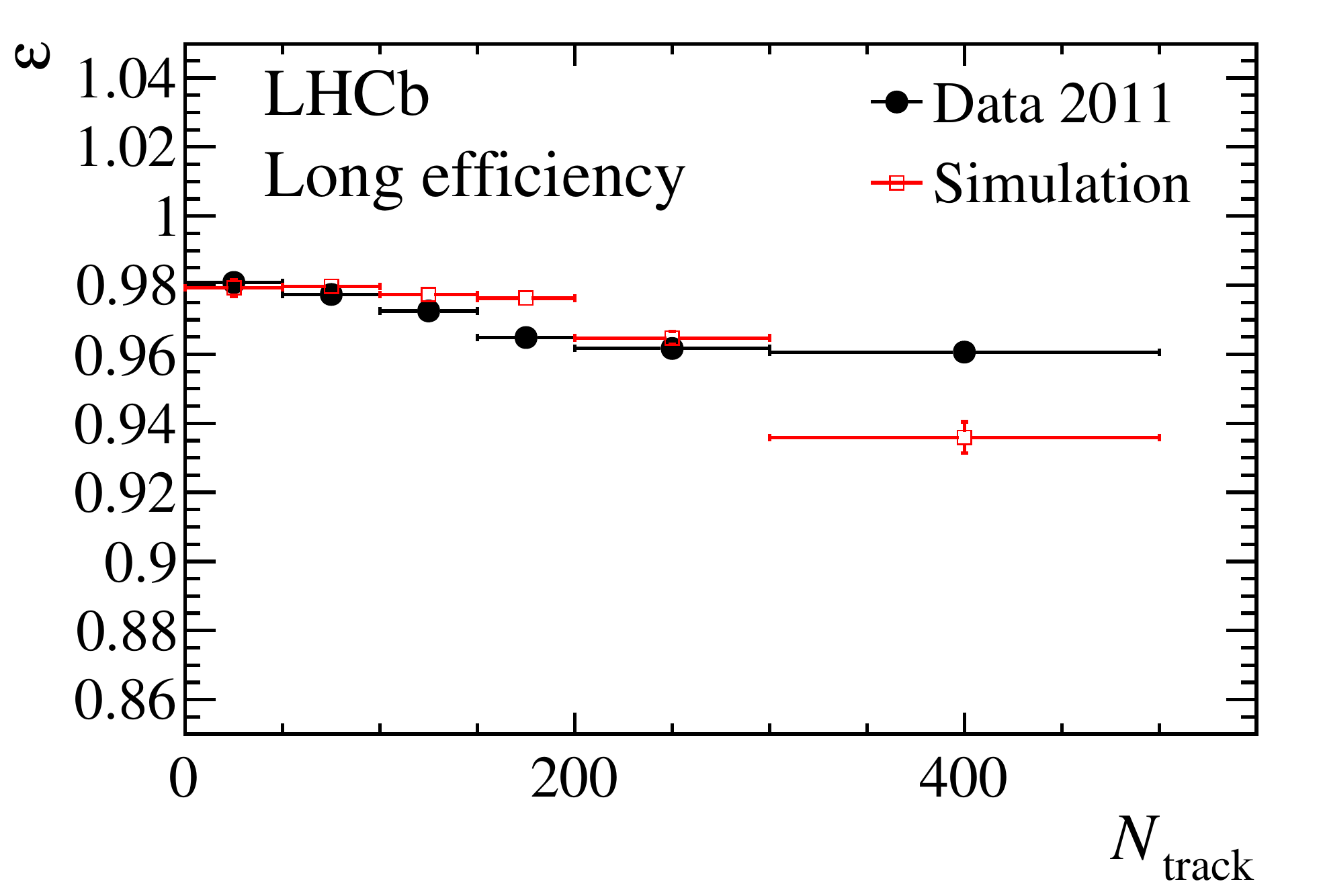}
    \includegraphics[width=0.46\textwidth]{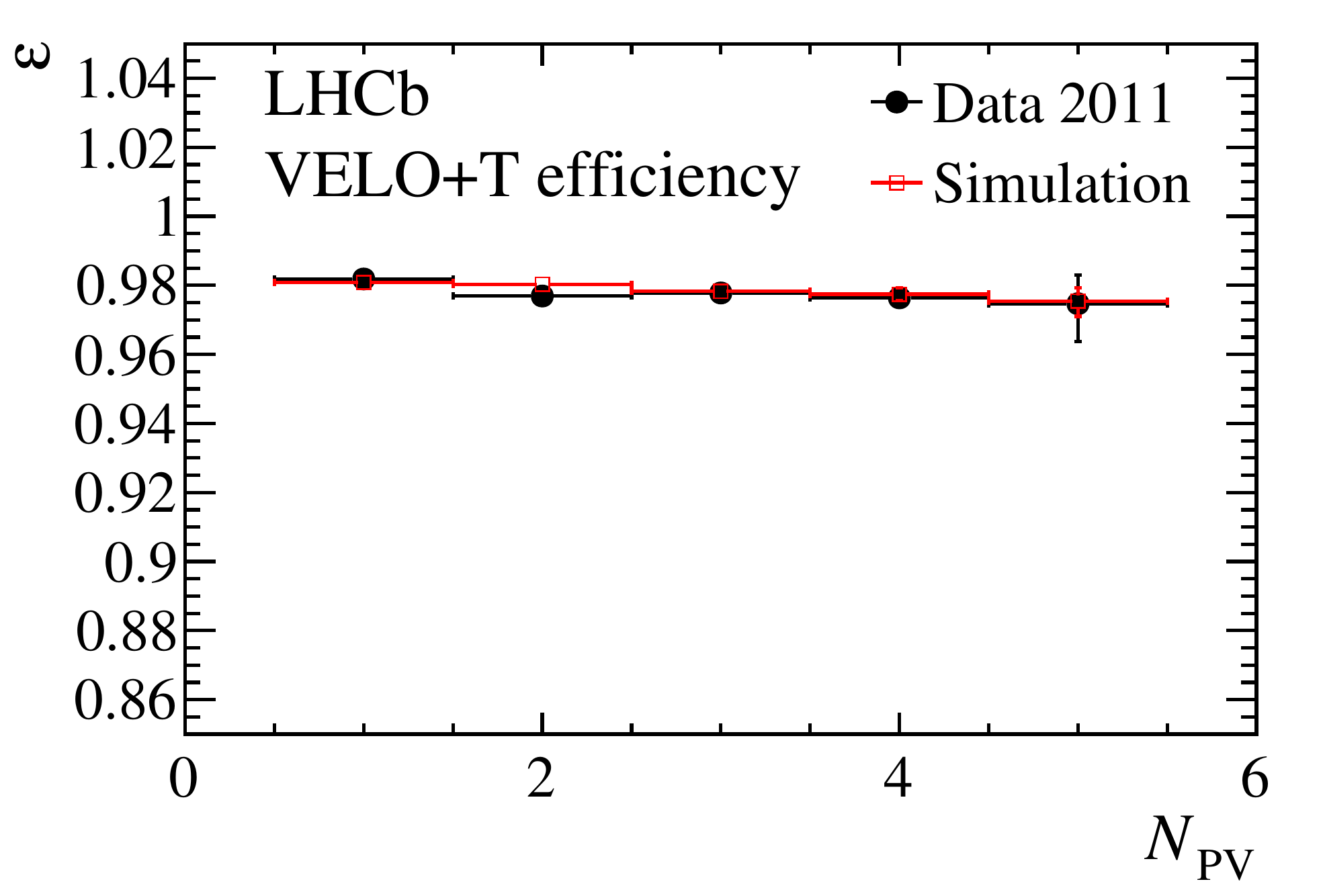}
    \includegraphics[width=0.46\textwidth]{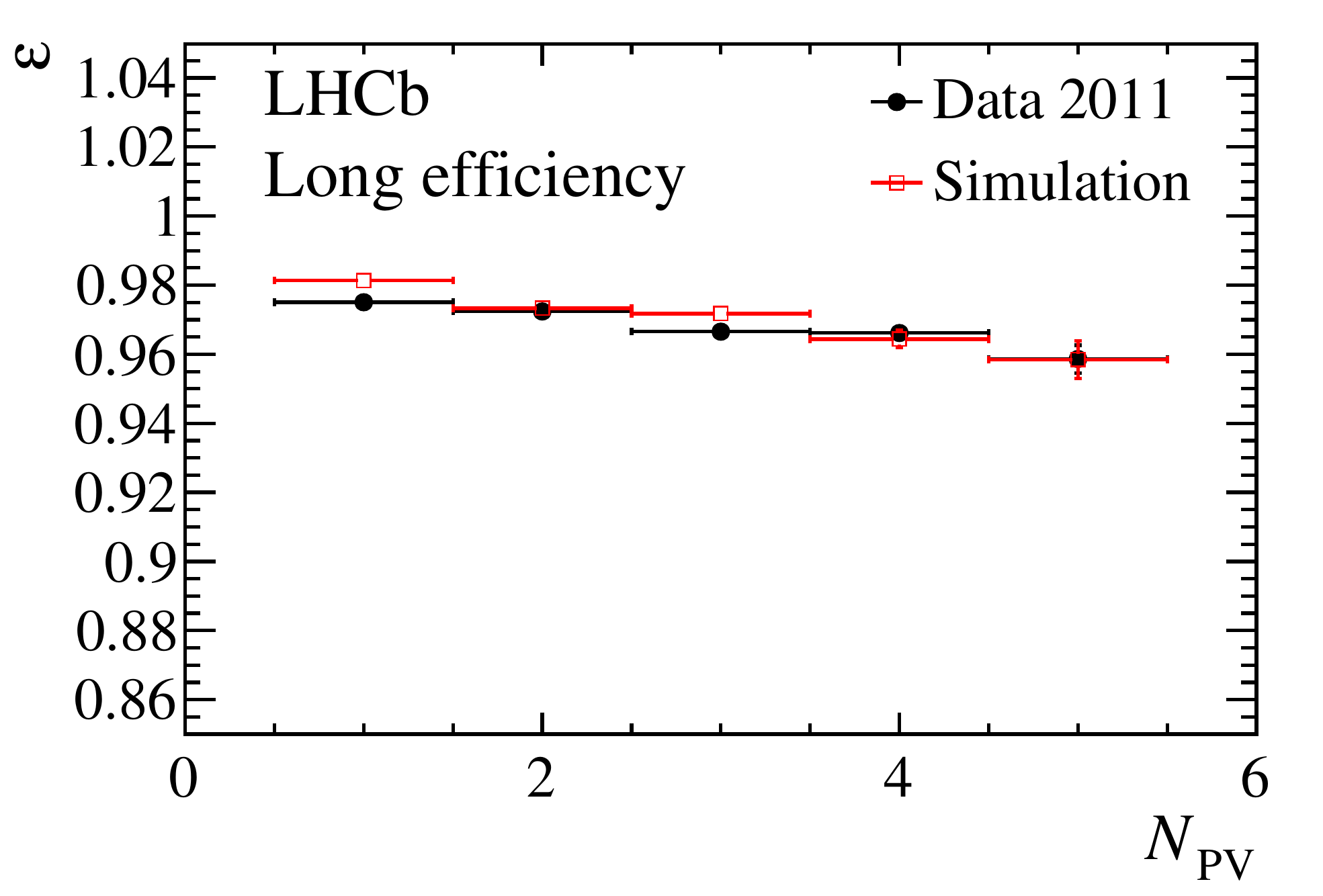}
  \end{center}
  \caption{\small Track reconstruction efficiencies for the 2011 data and for
    weighted simulation. The left-hand column shows the results of the combined
    method while the right-hand column shows the results of the long method. The
    efficiency is shown as a function of $p$ (first row),
     $\eta$ (second row), $N_{\rm track}$ (third row), and $N_{\rm PV}$ (fourth row). The error bars indicate the
  statistical uncertainties.}
  \label{fig:effLong2011}
\end{figure}

\begin{figure}
  \begin{center}
    \includegraphics[width=0.46\textwidth]{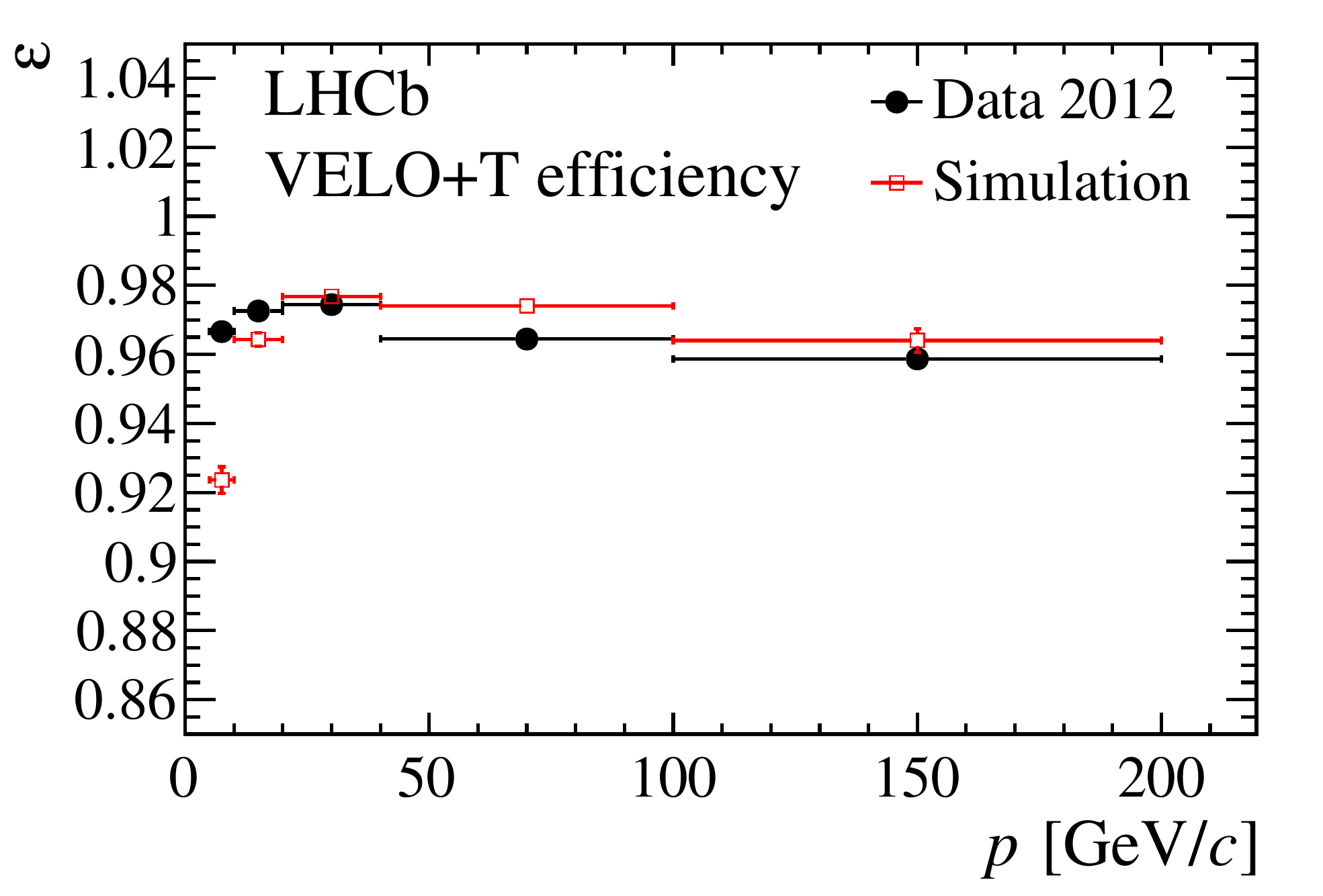}
    \includegraphics[width=0.46\textwidth]{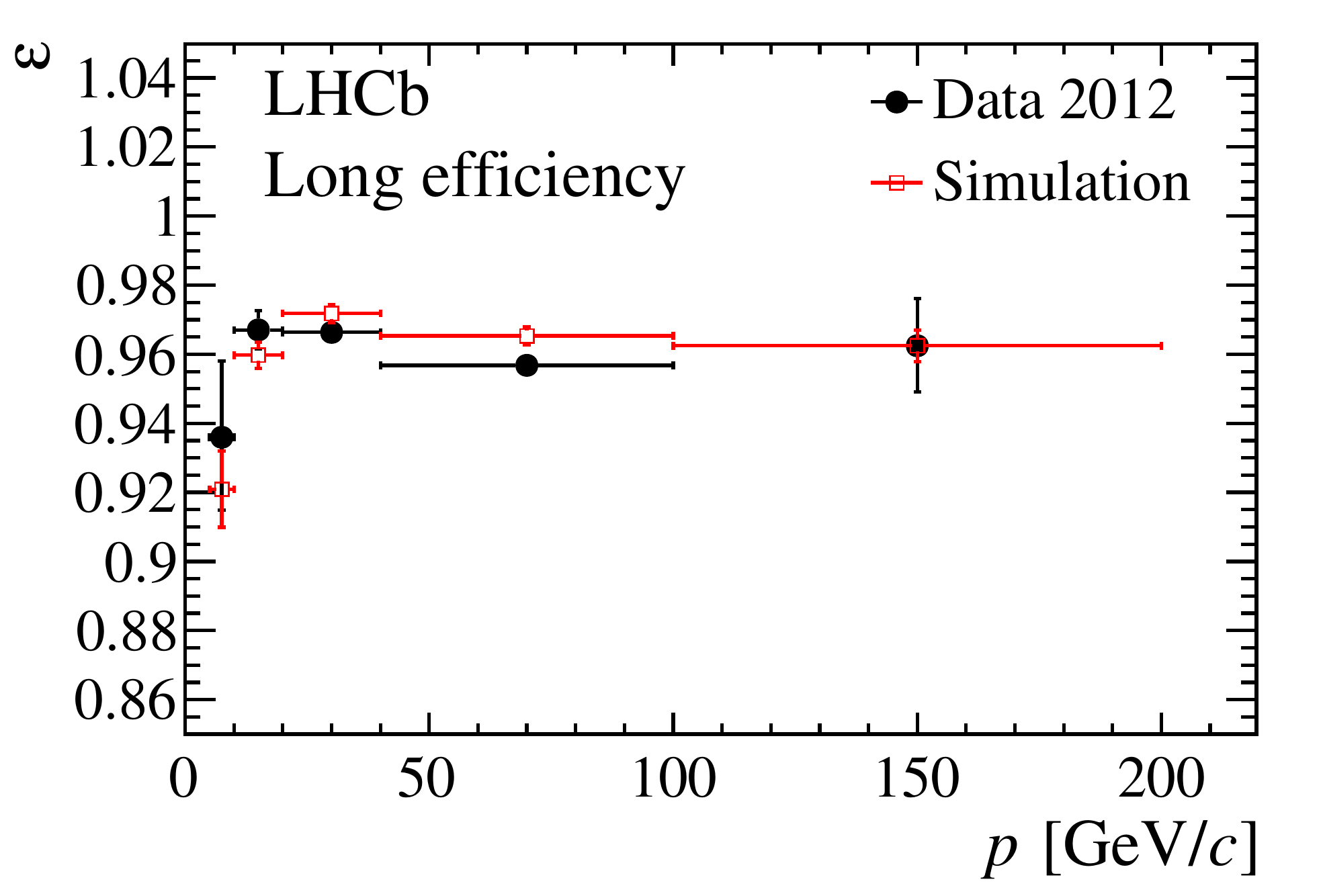}
    \includegraphics[width=0.46\textwidth]{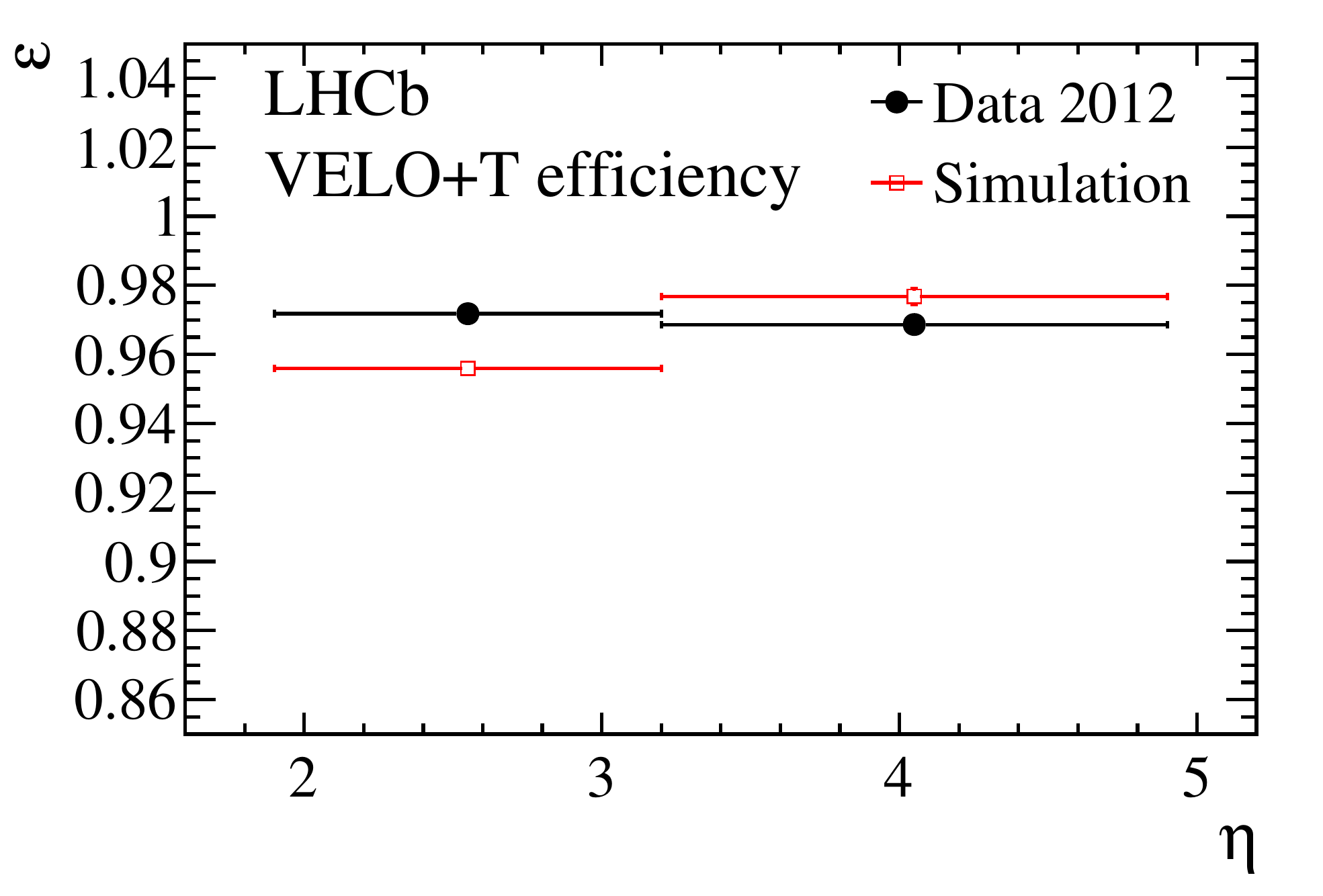}
    \includegraphics[width=0.46\textwidth]{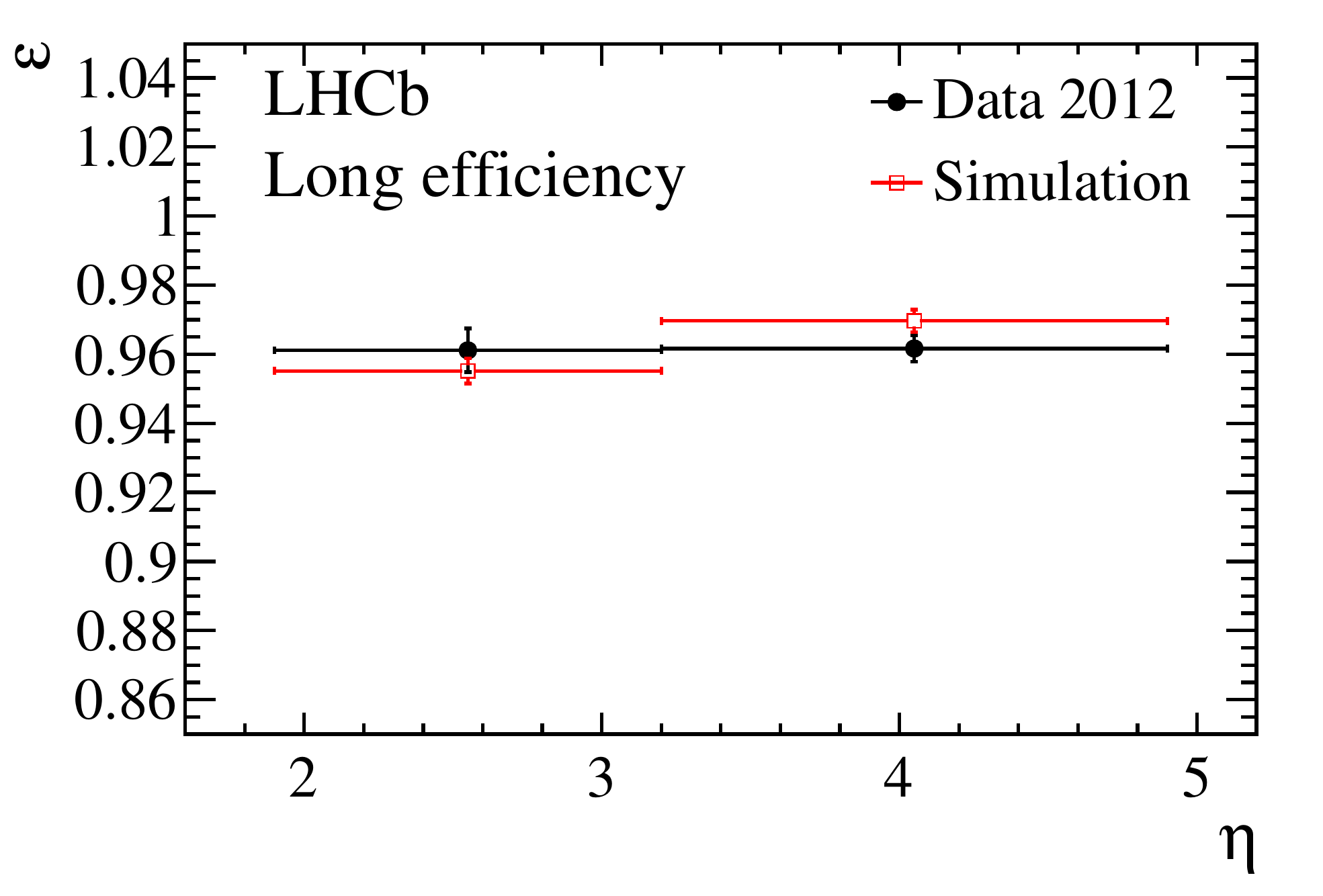}
    \includegraphics[width=0.46\textwidth]{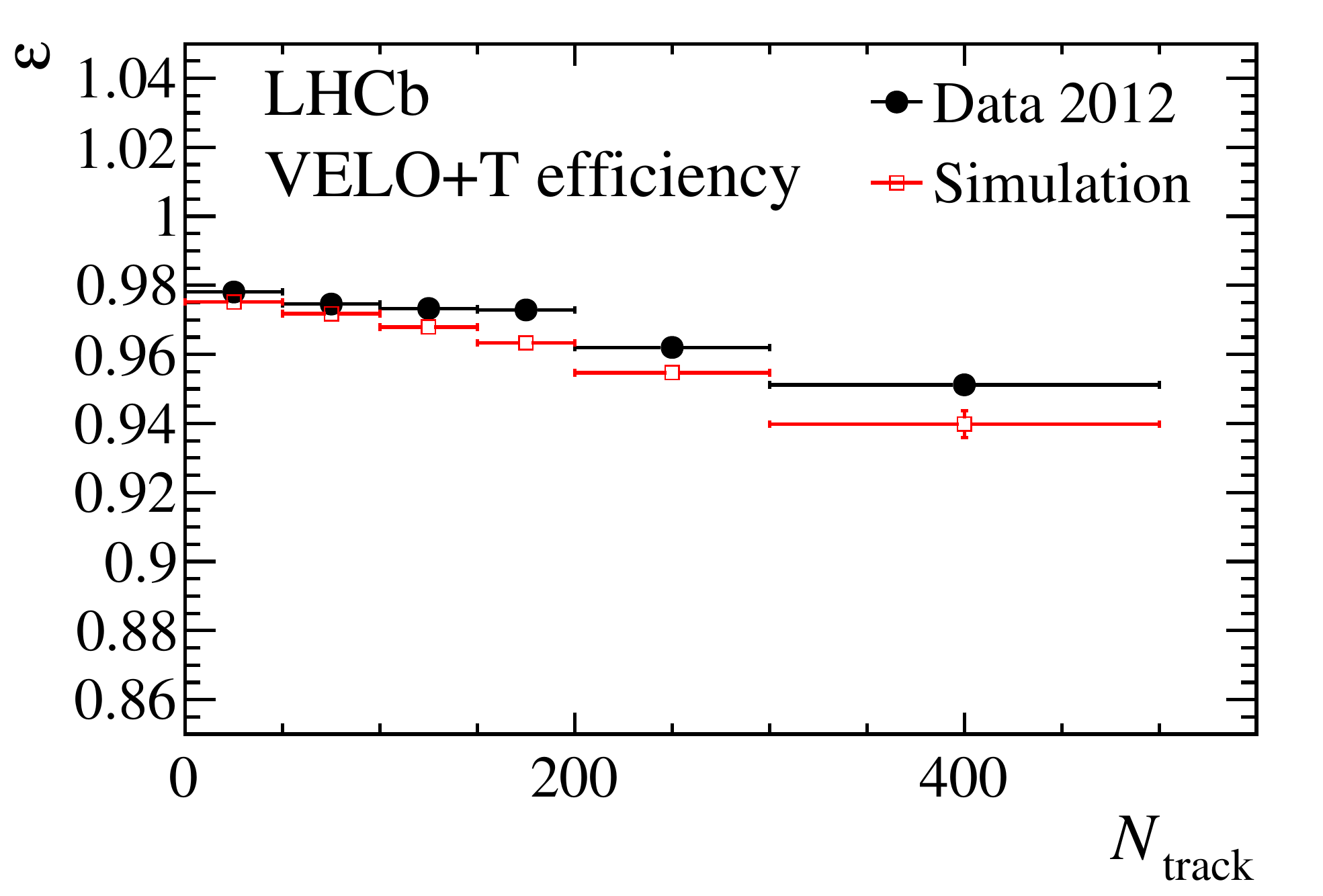}
    \includegraphics[width=0.46\textwidth]{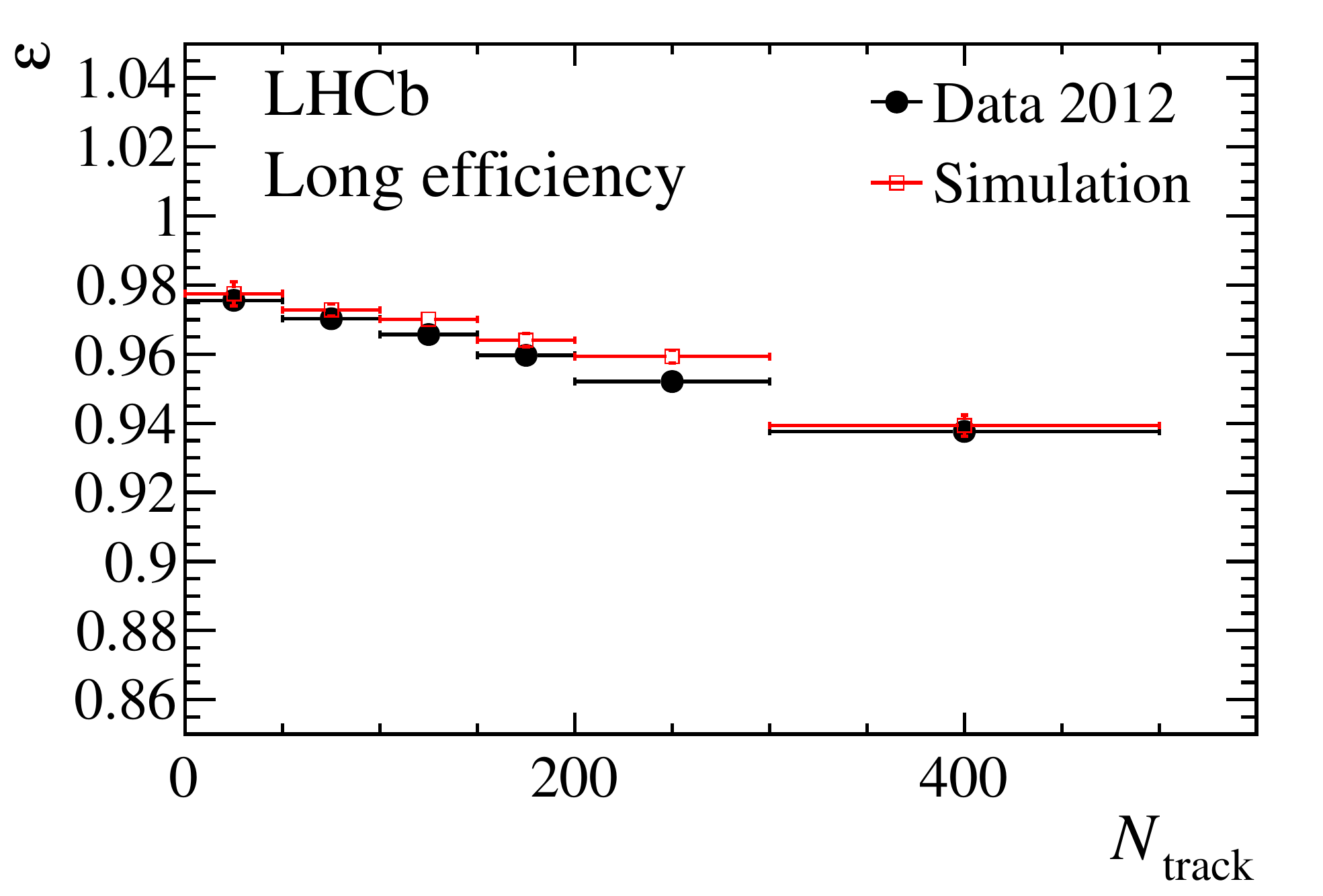}
    \includegraphics[width=0.46\textwidth]{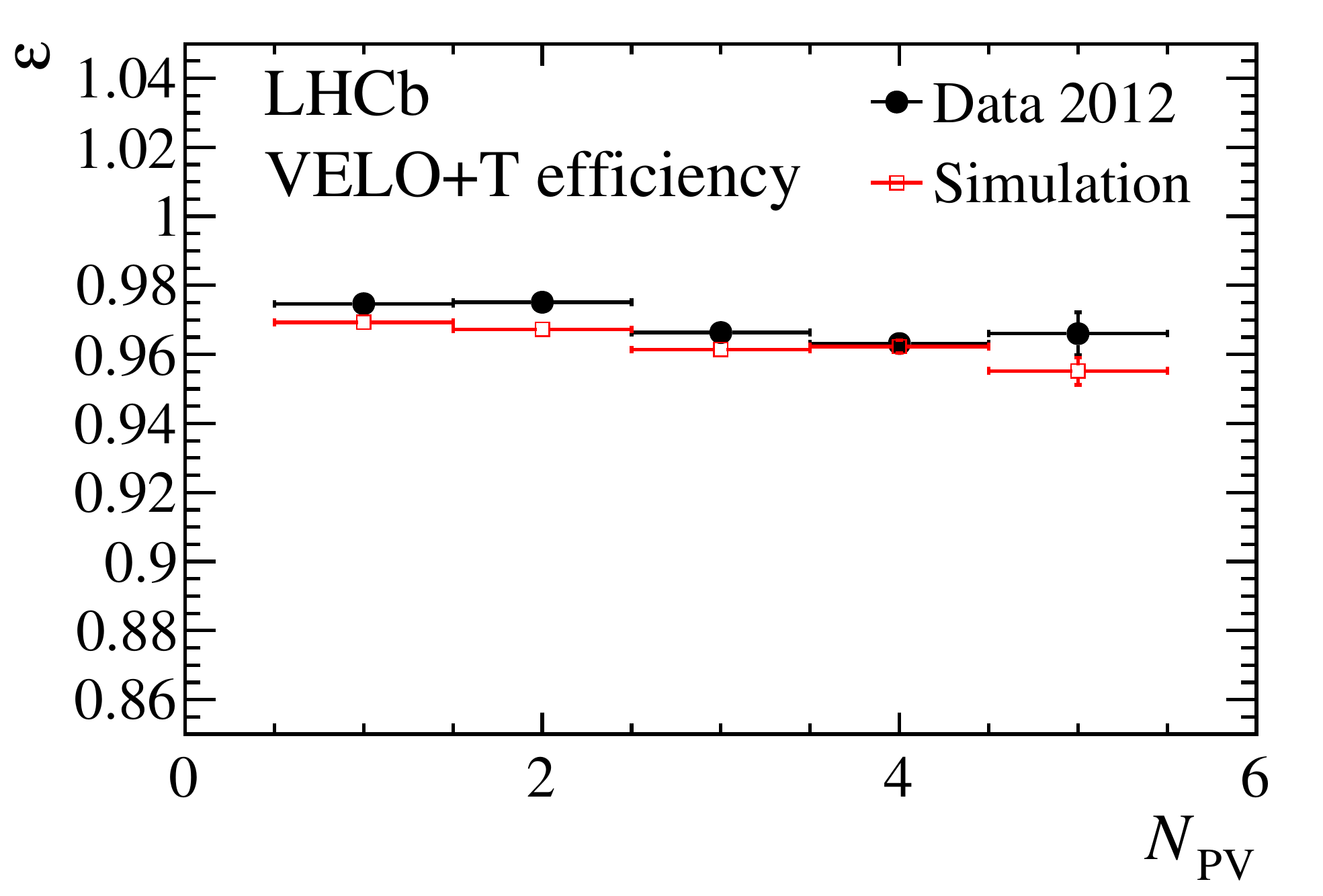}
    \includegraphics[width=0.46\textwidth]{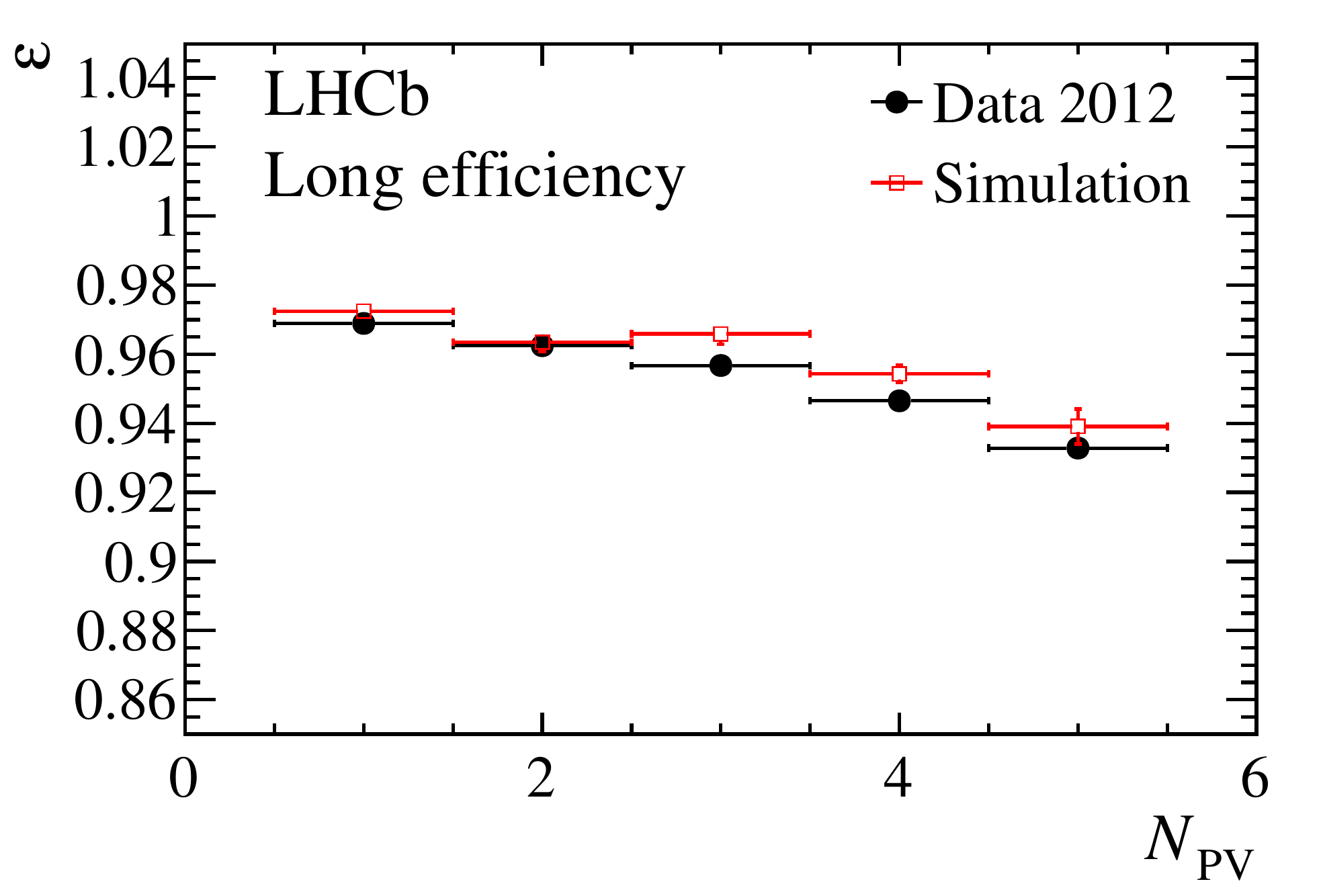}
  \end{center}
  \caption{\small Track reconstruction efficiencies for the 2012 data and for
    weighted simulation. The left-hand column shows the results of the combined
    method while the right-hand column shows the results of the long method. The
    efficiency is shown as a function of $p$ (first row),
     $\eta$ (second row), $N_{\rm track}$ (third row), and $N_{\rm PV}$ (fourth row). The error bars indicate the
  statistical uncertainties.}
  \label{fig:effLong2012}
\end{figure}

The efficiency to reconstruct long tracks mainly depends on the
particle kinematics and the number of charged particles in an event. As a parametrisation $p$,
$\eta$ and $N_\text{track}$ are chosen, as the track reconstruction efficiency shows the largest dependence on these three observables. 
The simulated events are weighted
according to the $N_\text{track}$ distribution observed in data. The track reconstruction efficiencies for the combination of the VELO and T-station methods and for the long method are shown for the different data-taking periods
in Figs.~\ref{fig:effLong2010}--\ref{fig:effLong2012} as a function of $p$,
$\eta$, $N_\text{track}$, and as a function of the number of reconstructed
primary vertices, $N_{\rm PV}$. The efficiency coming from the combination of the VELO and the T-station method is calculated by multiplying the individual efficiencies. Overall, a reasonable agreement is found between simulated and real data for all data-taking periods. As the agreement between the tag-and-probe based track reconstruction efficiency and the true track reconstruction efficiency (based on hit information) is within 1\%, the results shown in Figs.~\ref{fig:effLong2010}--\ref{fig:effLong2012} give an accurate description of the efficiency in simulation.

\addtolength{\abovecaptionskip} {-8mm}
\addtolength{\abovecaptionskip} {5mm}

\subsection{Efficiency ratios}

The efficiency ratio is defined as the efficiency measured in data
divided by the efficiency
measured in simulation,
\begin{equation}
  {\rm ratio}=\frac{\varepsilon_{\rm data}}{\varepsilon_{\rm sim}} \;\;\; .
\end{equation}
The efficiency dependence versus $N_{\rm track}$ and $N_{\rm PV}$ is reasonably well
described in the simulation, see Figs.~\ref{fig:effLong2010}--\ref{fig:effLong2012}: When fitting a first-order polynomial to the efficiency distributions in simulation and real data, the slopes agree with each other within 2 standard deviations, except for the efficiency as a function of the number of tracks in the combination of the VELO and T-station method in 2012. It is therefore sufficient to weight the simulated events to establish agreement in $N_{\rm track}$ while the efficiency ratio is determined in bins of $p$
and $\eta$. The number of bins is chosen to keep the statistical uncertainty
in each bin sufficiently small.
For the final result, the weighted average of the combined and long method is
taken in each bin of $p$ and $\eta$.

\begin{figure}
  \begin{center}
    \includegraphics[width=0.47\textwidth]{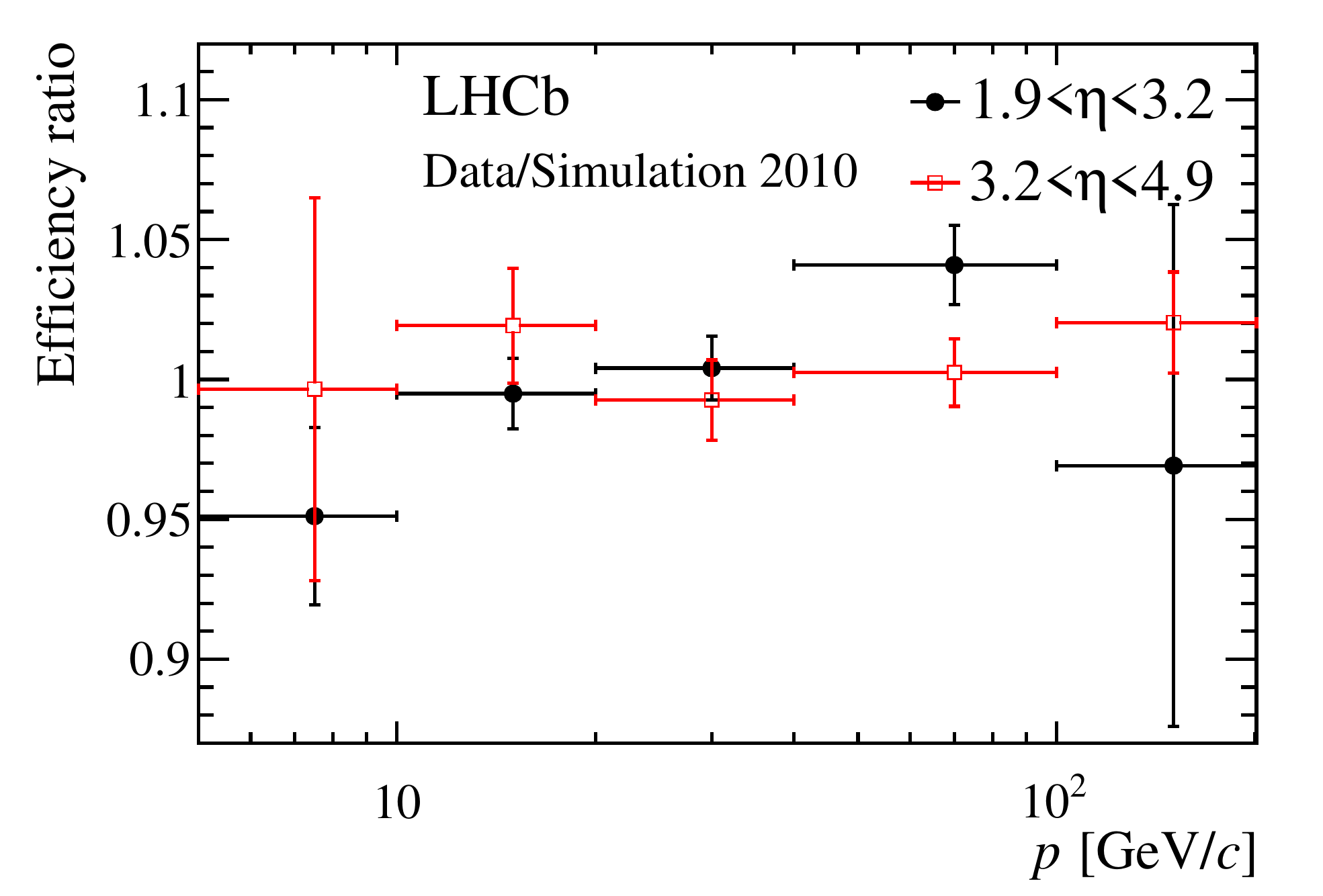}
    \includegraphics[width=0.47\textwidth]{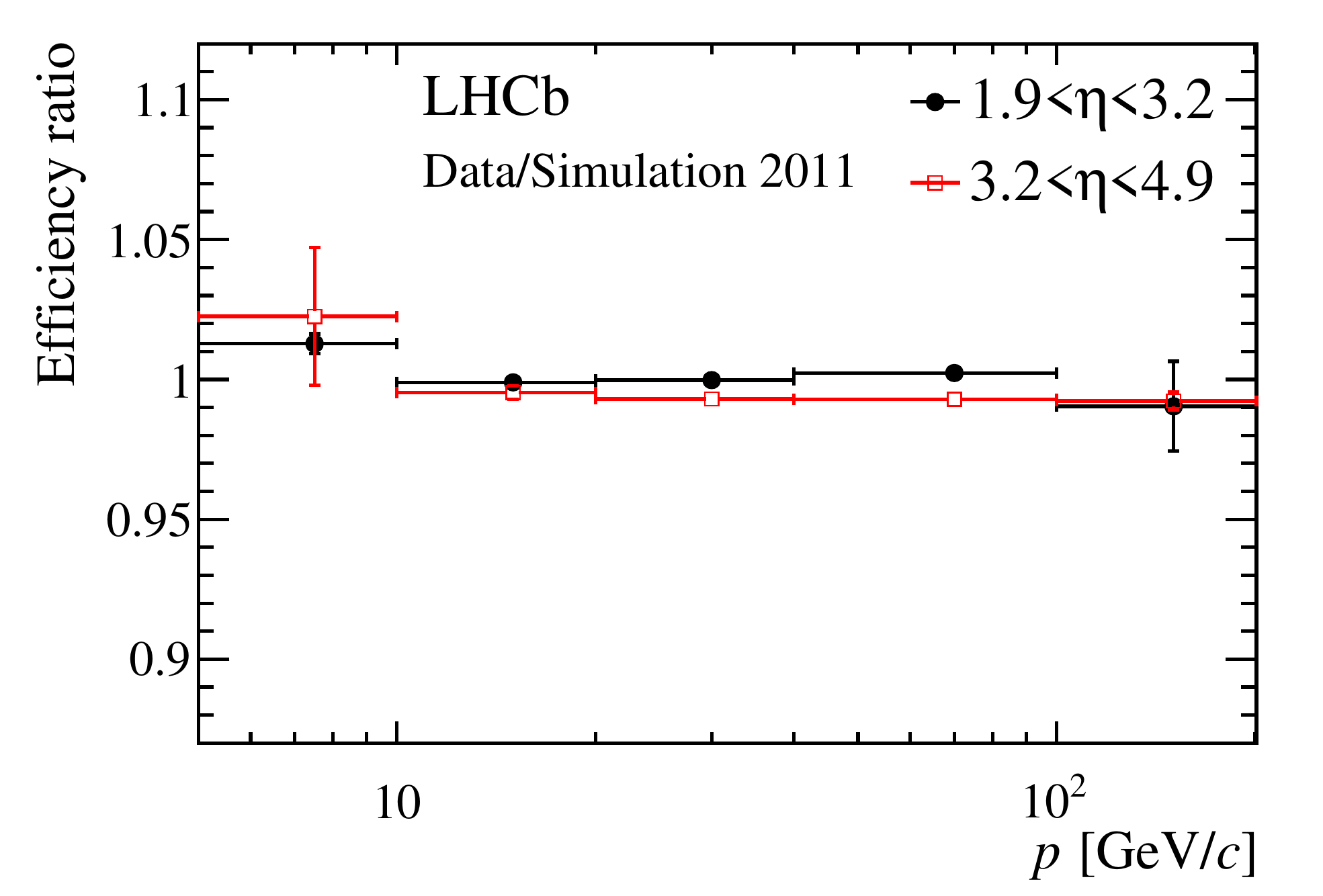}
    \includegraphics[width=0.47\textwidth]{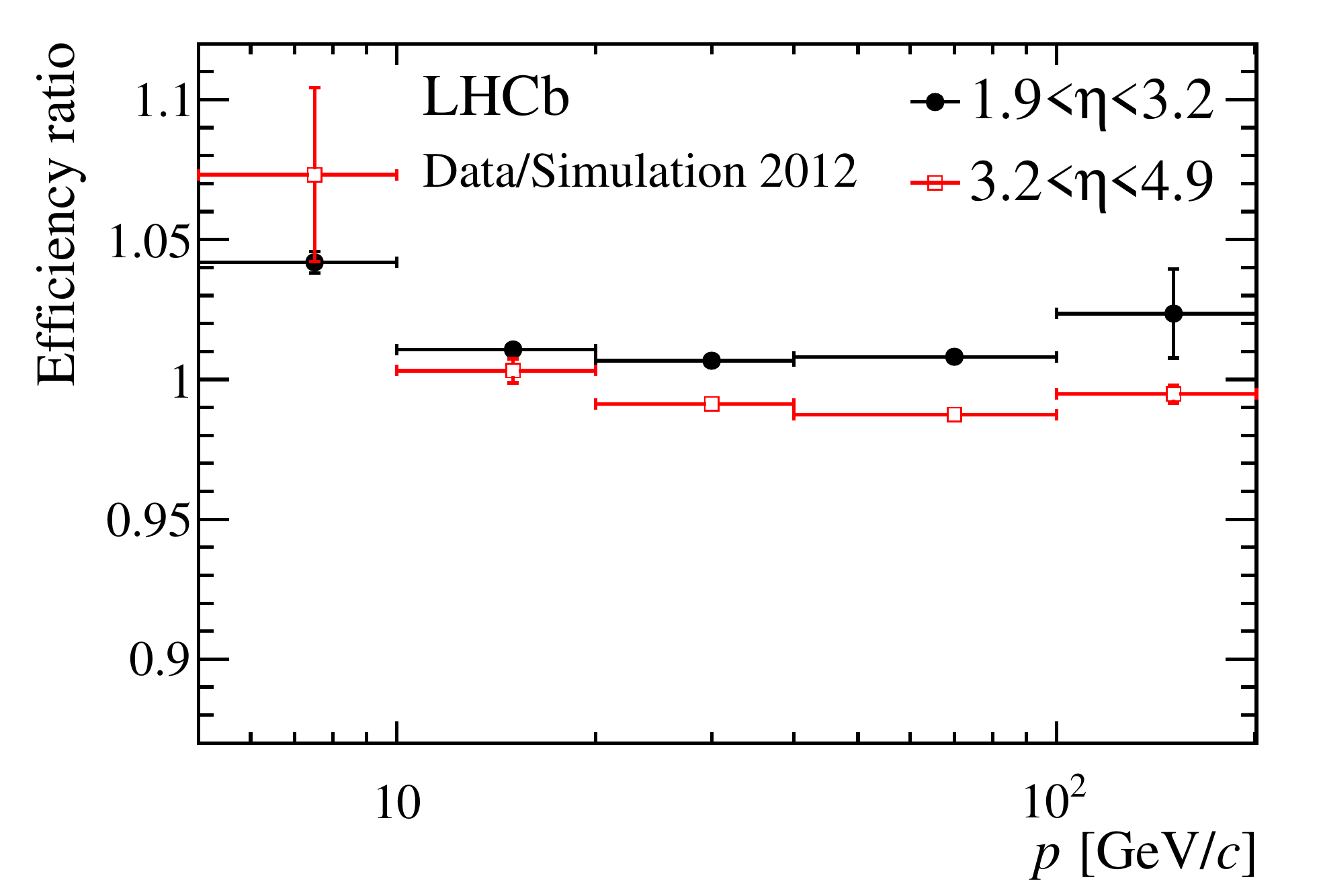}
  \end{center}
  \caption{Track reconstruction efficiency ratios as a function of $p$
  between data and simulation for (left) 2010 data, (right) 2011 data, and (bottom) 2012 data.}
  \label{fig:ratio}
\end{figure}

Figure~\ref{fig:ratio} shows the efficiency ratio versus $p$ for run I,
weighted by the event track multiplicity observed in data; the data are split
into two ranges of $\eta$.
Overall a good agreement of the track finding efficiency 
is found between events in simulation and in data for all data taking periods and 
most momenta and pseudorapidity regions. The difference between the track finding efficiencies 
is generally smaller than 1\% between events from simulation and data and no trend 
can be observed for the 2011 and 2012 dataset, with the number of events being 
too low to draw conclusions from the 2010 dataset. The agreement is worse for tracks with momentum below 10\gevc, which might point to a less accurate modelling of multiple scattering effects in the simulation.

The overall efficiency ratio and its uncertainty depend on the particle
distribution of the data in terms of $p$ and $\eta$. Using the momentum spectrum of the $\jpsi$ decay products obtained with the VELO method from data, an average efficiency ratio is found of
$0.994\pm0.007$ for 2010 data, $0.9983\pm0.0009$ for 2011 data and
$1.0053\pm0.0008$ for 2012 data. The uncertainties represent the statistical
uncertainties only. The ratio is close to one in all three cases as different features seen in the efficiency distributions in simulation and data average out when integrating over the full momentum spectrum or pseudorapidity range.

\section{Systematic uncertainties}
\label{sect:sysuncertainty}

Small differences in the ratio of efficiencies are seen when reweighting the
simulated samples in different parameters such as the number of primary
vertices, or the number of hits or tracks in the different subdetectors. The
largest of these differences is taken as a systematic uncertainty and amounts
to $0.4\%$. 
No systematic uncertainty is assigned for the agreement of the track reconstruction efficiency determined by the tag-and-probe method and the hit-based method (which is on the order of 1\%), as the differences cancel when forming the efficiency ratio. Accordingly, no systematic uncertainties are assigned for the fit model as these cancel when forming the fraction of reconstructed $\jpsi$ decays where the probe can be matched to a long track. It has been checked that this is true for a range of fit models, the largest variation being 0.2\%. Furthermore, no systematic uncertainty is assigned to the possible matching of a correctly reconstructed probe track to a fake long track, as the requirement for a large overlap in the subdetectors ensure that both reconstructed tracks are either real tracks or fake tracks, where the latter would not peak at the \jpsi mass. No systematic uncertainty is assigned for the fact that the VELO + T-station method and the long method show slightly different results in Figs.~\ref{fig:effLong2010}--\ref{fig:effLong2012}, as both methods probe different momentum spectra and any residual difference will cancel when forming the ratio with simulation. No systematic uncertainty is assigned for the double-counting of the matching efficiency in the combined method, as this efficiency is very close to 100\%, and any uncertainty would get further reduced when forming the ratio with simulation. No systematic uncertainty is assigned for the large difference for the VELO + T efficiency between simulation and data at low momenta in 2011 and 2012, as this is automatically taken into account when forming the ratio of efficiencies. Despite this difference, the integrated track reconstruction efficiencies between simulation and data are in agreement due to compensation of this effect for high momenta, where the efficiency is higher in simulation than in data.

\section{Hadronic interactions}
The methods presented in this paper are based on muons and require that they reach the muon stations. Thus, these methods are not sensitive to the
effects from hadronic interactions and large-angle scatterings with the
detector material. For hadrons, the largest effect is due to hadronic
interactions. The cross section depends on the particle type, charge and the
momentum. A simulation of $\Bz\to\jpsi\Kstarz$ decays (where
$\Kstarz\to\Kp\pim$) shows that about 11\% of the kaons (averaged over positive
and negative kaons) and about 14\% of the pions cannot be reconstructed due to
hadronic interactions that occur before the last T~station. This number depends
primarily on the momentum of the particle. Due to the uncertainty on the
material budget and consequently on the interaction with the detector material, the
reconstruction efficiency obtained from simulation has an intrinsic
uncertainty, which is not accounted for in the track reconstruction
efficiencies measured with muons. When assuming that the total material
budget in the simulation has an uncertainty of 10\%, the systematic uncertainty
due to hadronic interactions is between 1.1--1.4\%. The 10\% uncertainty is used as a conservative upper limit and is composed as follows: for the VELO a calculation in Ref.~\cite{VeloPerformance} shows an uncertainty on the material budget of 6\%. No direct measurements exist for the T and TT stations. However, weight measurements for the Inner Tracker for the silicon sensors and the detector boxes give an accuracy of 2\%, while an agreement of 5\% is reached for the cables and the support structure\cite{Fave:1134016, Fave:1134017}. The Outer Tracker modules have been weighted and this measurement is precise to about 1\% \cite{Nardulli:815493}. Furthermore, the sum of the weights of the individual components of a module adds up to the total weight of a module within the uncertainties. Taking into account that some level of detail is missing in the detector description in the simulation, an uncertainty of 5\% is assumed for the outer tracker. Weight measurements for the sensor modules and the insulation material of TT have been performed. Given the detail of the detector description\cite{Salzmann:1140703} an uncertainty of 5\% on the material budget is well justified. The beam-pipe was implemented in the software following the design drawings, where a precision better than 10\% for all pieces was confirmed following measurements after production. The solid radiator (aerogel) and the gas radiator (\cfourften) contribute more than two-third of the material budget for the RICH1 detector \cite{Brook:897981}. The amount of aerogel is known up to 2\% and the differences between 2011 and 2012 are accounted for in the simulation. The density of the \cfourften was monitored, with the RMS of the distribution being about 1\%. The other components of RICH1 have a smaller contribution to the interaction length. 
The overall uncertainty of 10\% for the full material budget was then chosen to also take uncertainties on the \geant cross-sections and additional uncertainties, coming from simplified descriptions of the detector elements in the simulation, into account.

\section{Conclusion}
\label{sec:results}

Track reconstruction efficiencies at LHCb have been measured using a
tag-and-probe method with $\decay{\jpsi}{\mup\mun}$ decays. The average
efficiency is better than $95\%$ in the momentum region $5\gevc < p < 200\gevc$ and in the pseudorapidity region $2 < \eta < 5$, 
which covers the phase space of LHCb. The
uncertainty per track is below $0.5\%$ for muons and below 1.5\% for pions and kaons, where the larger uncertainty takes the uncertainty on hadronic interactions into account. All uncertainties have been added in quadrature. Furthermore, the ratio of the track reconstruction efficiency of muons in data and simulation is measured, where an uncertainty of  $0.8\,\%$ for data collected in 2010 and an uncertainty of $0.4\,\%$ for data collected in 2011 and 2012 is achieved. The integrated efficiency ratios for all three years of data taking are compatible with unity. This result presents a significant improvement over the uncertainties determined with
previous methods ranging from $3$ to $4\%$.

\section*{Acknowledgements}

\noindent We express our gratitude to our colleagues in the CERN
accelerator departments for the excellent performance of the LHC. We
thank the technical and administrative staff at the LHCb
institutes. We acknowledge support from CERN and from the national
agencies: CAPES, CNPq, FAPERJ, and FINEP (Brazil); NSFC (China);
CNRS/IN2P3 (France); BMBF, DFG, HGF, and MPG (Germany); SFI (Ireland); INFN (Italy); 
FOM and NWO (The Netherlands); MNiSW and NCN (Poland); MEN/IFA (Romania); 
MinES and FANO (Russia); MinECo (Spain); SNSF and SER (Switzerland); 
NASU (Ukraine); STFC (United Kingdom); NSF (USA).
The Tier1 computing centres are supported by IN2P3 (France), KIT and BMBF 
(Germany), INFN (Italy), NWO and SURF (The Netherlands), PIC (Spain), GridPP 
(United Kingdom).
We are indebted to the communities behind the multiple open 
source software packages on which we depend. We are also thankful for the 
computing resources and the access to software R\&D tools provided by Yandex LLC (Russia).
Individual groups or members have received support from 
EPLANET, Marie Sk\l{}odowska-Curie Actions, and ERC (European Union), 
Conseil g\'{e}n\'{e}ral de Haute-Savoie, Labex ENIGMASS, and OCEVU, 
R\'{e}gion Auvergne (France), RFBR (Russia), XuntaGal, and GENCAT (Spain), Royal Society and Royal
Commission for the Exhibition of 1851 (United Kingdom).

\addcontentsline{toc}{section}{References}
\bibliographystyle{LHCb}
\bibliography{main,LHCb-PAPER,LHCb-CONF,LHCb-DP,LHCb-TDR}

\end{document}